\documentclass[10pt,english,twocolumn,aps,prl,superscriptaddress]{revtex4-1}
\usepackage[T1]{fontenc}
\usepackage[latin9]{inputenc}
\usepackage{amsmath}
\usepackage{amssymb}
\usepackage{graphicx}
\usepackage[pdftex,colorlinks=true,urlcolor= blue,linkcolor=Blue,citecolor=RedViolet]{hyperref}
\usepackage{hyperref}
\usepackage{soul}
\usepackage[usenames, dvipsnames]{color} 
\usepackage[subrefformat=parens,labelformat=parens,caption=false]{subfig}
\makeatletter
\usepackage{babel}

\usepackage{refstyle}

\makeatother

\begin{document}

\global\long\def\ket#1{|#1\rangle}

\global\long\def\Ket#1{\left|#1\right>}

\global\long\def\bra#1{\langle#1|}

\global\long\def\Bra#1{\left<#1\right|}

\global\long\def\bk#1#2{\langle#1|#2\rangle}

\global\long\def\BK#1#2{\left\langle #1\middle|#2\right\rangle }

\global\long\def\kb#1#2{\ket{#1}\!\bra{#2}}

\global\long\def\KB#1#2{\Ket{#1}\!\Bra{#2}}

\global\long\def\mel#1#2#3{\bra{#1}#2\ket{#3}}

\global\long\def\MEL#1#2#3{\Bra{#1}#2\Ket{#3}}

\global\long\def\n#1{|#1|}

\global\long\def\N#1{\left|#1\right|}

\global\long\def\ns#1{|#1|^{2}}

\global\long\def\NS#1{\left|#1\right|^{2}}

\global\long\def\nn#1{\lVert#1\rVert}

\global\long\def\NN#1{\left\lVert #1\right\rVert }

\global\long\def\nns#1{\lVert#1\rVert^{2}}

\global\long\def\NNS#1{\left\lVert #1\right\rVert ^{2}}

\global\long\def\ev#1{\langle#1\rangle}

\global\long\def\EV#1{\left\langle #1\right\rangle }

 \global\long\def\ha{\hat{a}}

\global\long\def\hb{\hat{b}}

\global\long\def\hc{\hat{c}}

\global\long\def\hd{\hat{d}}

\global\long\def\he{\hat{e}}

\global\long\def\hf{\hat{f}}

\global\long\def\hg{\hat{g}}

\global\long\def\hh{\hat{h}}

\global\long\def\hi{\hat{i}}

\global\long\def\hj{\hat{j}}

\global\long\def\hk{\hat{k}}

\global\long\def\hl{\hat{l}}

\global\long\def\hm{\hat{m}}

\global\long\def\hn{\hat{n}}

\global\long\def\ho{\hat{o}}

\global\long\def\hp{\hat{p}}

\global\long\def\hq{\hat{q}}

\global\long\def\hr{\hat{r}}

\global\long\def\hs{\hat{s}}

\global\long\def\hu{\hat{u}}

\global\long\def\hv{\hat{v}}

\global\long\def\hw{\hat{w}}

\global\long\def\hx{\hat{x}}

\global\long\def\hy{\hat{y}}

\global\long\def\hz{\hat{z}}

\global\long\def\hA{\hat{A}}

\global\long\def\hB{\hat{B}}

\global\long\def\hC{\hat{C}}

\global\long\def\hD{\hat{D}}

\global\long\def\hE{\hat{E}}

\global\long\def\hF{\hat{F}}

\global\long\def\hG{\hat{G}}

\global\long\def\hH{\hat{H}}

\global\long\def\hI{\hat{I}}

\global\long\def\hJ{\hat{J}}

\global\long\def\hK{\hat{K}}

\global\long\def\hL{\hat{L}}

\global\long\def\hM{\hat{M}}

\global\long\def\hN{\hat{N}}

\global\long\def\hO{\hat{O}}

\global\long\def\hP{\hat{P}}

\global\long\def\hQ{\hat{Q}}

\global\long\def\hR{\hat{R}}

\global\long\def\hS{\hat{S}}

\global\long\def\hT{\hat{T}}

\global\long\def\hU{\hat{U}}

\global\long\def\hV{\hat{V}}

\global\long\def\hW{\hat{W}}

\global\long\def\hX{\hat{X}}

\global\long\def\hY{\hat{Y}}

\global\long\def\hZ{\hat{Z}}

\global\long\def\hap{\hat{\alpha}}

\global\long\def\hbt{\hat{\beta}}

\global\long\def\hgm{\hat{\gamma}}

\global\long\def\hGm{\hat{\Gamma}}

\global\long\def\hdt{\hat{\delta}}

\global\long\def\hDt{\hat{\Delta}}

\global\long\def\hep{\hat{\epsilon}}

\global\long\def\hvep{\hat{\varepsilon}}

\global\long\def\hzt{\hat{\zeta}}

\global\long\def\het{\hat{\eta}}

\global\long\def\hth{\hat{\theta}}

\global\long\def\hvth{\hat{\vartheta}}

\global\long\def\hTh{\hat{\Theta}}

\global\long\def\hio{\hat{\iota}}

\global\long\def\hkp{\hat{\kappa}}

\global\long\def\hld{\hat{\lambda}}

\global\long\def\hLd{\hat{\Lambda}}

\global\long\def\hmu{\hat{\mu}}

\global\long\def\hnu{\hat{\nu}}

\global\long\def\hxi{\hat{\xi}}

\global\long\def\hXi{\hat{\Xi}}

\global\long\def\hpi{\hat{\pi}}

\global\long\def\hPi{\hat{\Pi}}

\global\long\def\hrh{\hat{\rho}}

\global\long\def\hvrh{\hat{\varrho}}

\global\long\def\hsg{\hat{\sigma}}

\global\long\def\hSg{\hat{\Sigma}}

\global\long\def\hta{\hat{\tau}}

\global\long\def\hup{\hat{\upsilon}}

\global\long\def\hUp{\hat{\Upsilon}}

\global\long\def\hph{\hat{\phi}}

\global\long\def\hvph{\hat{\varphi}}

\global\long\def\hPh{\hat{\Phi}}

\global\long\def\hch{\hat{\chi}}

\global\long\def\hps{\hat{\psi}}

\global\long\def\hPs{\hat{\Psi}}

\global\long\def\hom{\hat{\omega}}

\global\long\def\hOm{\hat{\Omega}}

\global\long\def\hdgg#1{\hat{#1}^{\dagger}}

\global\long\def\cjg#1{#1^{*}}

\global\long\def\hsgx{\hat{\sigma}_{x}}

\global\long\def\hsgy{\hat{\sigma}_{y}}

\global\long\def\hsgz{\hat{\sigma}_{z}}

\global\long\def\hsgp{\hat{\sigma}_{+}}

\global\long\def\hsgm{\hat{\sigma}_{-}}

\global\long\def\hsgpm{\hat{\sigma}_{\pm}}

\global\long\def\hsgmp{\hat{\sigma}_{\mp}}

\global\long\def\dert#1{\frac{d}{dt}#1}

\global\long\def\dertt#1{\frac{d#1}{dt}}

\global\long\def\Tr{\text{Tr}}

\title{Inducing nontrivial qubit coherence through a controlled dispersive environment}

\author{Wallace S. Teixeira}
\affiliation{Centro de Ciências Naturais e Humanas, Universidade Federal do ABC, Santo André, 09210-170 São Paulo, Brazil}
\author{Fernando Nicacio}
\affiliation{Instituto de Física, Universidade Federal do Rio de Janeiro, Caixa Postal 68528, Rio de Janeiro, RJ 21941-972, Brazil}
\author{Fernando L. Semião}
\affiliation{Centro de Ciências Naturais e Humanas, Universidade Federal do ABC, Santo André, 09210-170 São Paulo, Brazil}

\begin{abstract}
We show how the dispersive regime of the Jaynes-Cummings model may serve as a valuable tool to the study of open quantum systems. We employ it in a bottom-up approach to build an environment that preserves qubit energy and induces varied coherence dynamics. We then present the derivation of a compact expression for the qubit coherence, applied here to the case of a finite number of thermally populated modes in the environment. We also discuss how the model parameters can be adjusted to facilitate the production of short-time monotonic decay (STMD) of the qubit coherence. Our results provide a broadly applicable platform for the investigation of energy-conserving open system dynamics which is fully within the grasp of current quantum technologies.
\end{abstract}

\maketitle

\section{Introduction}

Studying and understanding the role of decoherence in open quantum systems has been a major topic in quantum technology. At the same time that decoherence is harmful to quantum information by washing out superposition aspects of quantum states \cite{Schlosshauer2008}, it can also be helpful in other tasks such as energy transport in quantum networks \cite{Plenio2008,Sinayskiy2012,Marais2013}. In other scenarios, it might be desirable to engineer it for multiple applications \cite{Poyatos1996,Carvalho2001,Roszak2015}. Several experiments have also unveiled the essential aspects of decoherence in controlled quantum systems \cite{Brune1996,Myatt2000,Bertet2005,Buono2012,Schneider2014}. In a more fundamental level, decoherence is expected to be involved in the emergence of the classical world from within the set of  quantum rules \cite{Zurek1991}.

In the simplest case, decoherence of a two-level system (qubit) follows from its linear coupling to a thermal reservoir consisting of a collection of an infinite number of independent (noninteracting) quantum harmonic oscillators \cite{Leggett1987} or two-level systems \cite{Proko2000}. In such descriptions, a lack of control and accessibility to the degrees of freedom of the environment is assumed. In this work, we propose the study of qubit pure dephasing, which is a  form of decoherence, in a fully controllable environment built \textsl{from the bottom up}. In other words, we blend together the advances in controlled quantum systems and open systems theory to investigate qubit decoherence in a fully controlled and finite environment whose number of degrees of freedom can be carefully increased. Other approaches to the engineering of pure dephasing have been proposed for the harmonic motion of a trapped ion \cite{Turchette2000} and, more recently, for the polarization of a photon in an environment composed by its frequency degree of freedom \cite{Liu2018}.

\begin{figure}
\includegraphics[width=0.9\linewidth]{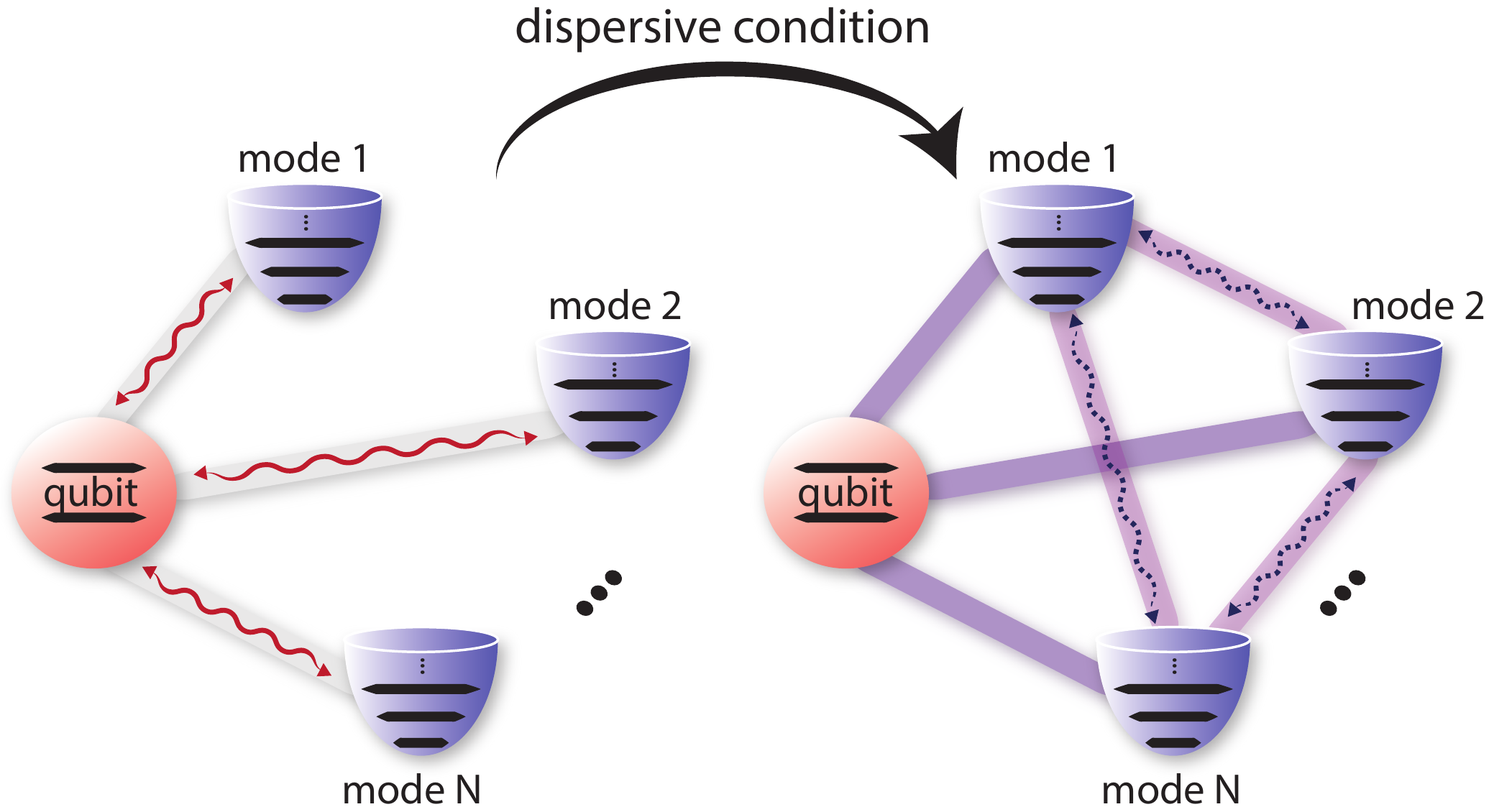} \caption{(Color online) Pictorial representation for the extended dispersive regime. Each mode corresponds to a distinct single-mode resonator.}
\label{fig:model} 
\end{figure}

Our approach is based on the multimode version of the dispersive regime of the Jaynes-Cummings model \cite{Jaynes1963}, 
where a qubit and a single mode of the electromagnetic field are  considerably out of resonance, preventing transitions between energy states of the free Hamiltonians. The qubit-mode coupling in this regime manifests itself through induced energy shifts in such energy levels. The dispersive limit of the Jaynes-Cummings model has been employed in a myriad of tasks. Important examples include the generation of superpositions of coherent states of opposite phases (``Schrödinger's cat'' states), nondemolition measurements in cavity QED \cite{Brune1992}, and more recently qubit readout in circuit QED \cite{Blais2004}, just to name a few. However, much less attention has been given for its use in the context of open quantum systems. This is precisely the proposal we put forward in this work: an environment consisting of $N$ modes dispersively coupled to the qubit, as depicted in Fig.~\ref{fig:model}. Interesting enough, the extension of the dispersive condition to $N$ modes induces a structure in the environment which now consists of coupled modes in contrast to the canonical models of decoherence mentioned before. The interplay between structure, frequencies, number of modes, and temperature promotes a very rich scenario where energy-conserving non-Markovian dynamics can be studied and applied, for instance, to the production of short-time monotonic decay (STMD) of the qubit coherence. In particular, given the lack of energy transitions which is inherit to the model, the dispersive qubit-mode interaction might serve as a building block for a qubit dephasing model with the distinct advantage of being fully controllable in several setups, as given evidence by the aforementioned applications.

\section{Model}

Let us consider a single qubit interacting with $N$ bosonic modes. Under the dipole and rotating-wave approximations, the total Hamiltonian of the system is thus described by the extended (multimode)
Jaynes-Cummings model ($\hbar=1$) \cite{Carmichael2013}: 
\begin{equation}
\hH=\frac{\omega_{0}}{2}\hsgz+\sum_{j=1}^{N}\omega_{j}\hdgg a_{j}\ha_{j}+\sum_{j=1}^{N}g_{j}\left(\hsgp\ha_{j}+\hsgm\hdgg a_{j}\right),\label{eq:HJCN}
\end{equation}
where $\omega_{0}$ is the frequency of the qubit, $\hsg_{i}$ ($i=x,y,z$)
are the Pauli matrices, $\omega_{j}$ is the frequency of the $j$-th
mode described by the annihilation operator $\ha_{j}$, and $g_{j}$
is the coupling constant.
The operators $\hsgpm=\frac{1}{2}\left(\hsgx\pm i\hsgy\right)$ are
the ladder operators for the qubit. 
For the case of an environment being composed of a continuum of electromagnetic modes, Hamiltonian~(\ref{eq:HJCN}) has been exhaustively used to model \textsl{dissipative} qubit dynamics or spontaneous emission \cite{Carmichael2013,Leandro2009}. This problem has been tackled with perturbative \cite{Carmichael2013,Breuer1999,Vacchini2010}
and nonperturbative methods \cite{Garraway1997}, as well as subjected to Markov approximation \cite{Carmichael2013}
and slightly modified to accommodate structured environments \cite{Garraway1997,Mazzola2009}. These approaches have in common the fact that the environment formed by the modes has a frequency distribution which is essentially centered around the qubit frequency. In this way, whenever the resonant or quasi-resonant Hamiltonian ~(\ref{eq:HJCN}) is employed to build a reservoir model, the result is dissipative dynamics.
In the present work, however, we take a different route which aims at producing a nondissipative open system dynamics, i.e., pure dephasing with finite $N$. In order to do it, we will consider that the modes are far from resonance with the qubit and then take the dispersive limit of  Hamiltonian~(\ref{eq:HJCN}). Not only that, we take full advantage of the current high level of experimental control over dispersive interactions to propose a bottom-up approach. We provide analytical results for the coherence dynamics in our model.

In the interaction picture 
with respect to the free part of Hamiltonian (\ref{eq:HJCN}), 
the dynamics follows from 
\begin{equation}
\hH^{I}(t)=\sum_{j=1}^{N}g_{j}\left(\hsgp\ha_{j}e^{i\Delta_{j}t}+\text{\ensuremath{\hsgm\hdgg a_{j}e^{-i\Delta_{j}t}}}\right),\label{eq:HJCI}
\end{equation}
with $\Delta_{j}=\omega_{0}-\omega_{j}$ being the detuning between the qubit and mode $j$. The requirement
\begin{equation}
\N{\frac{g_{j}}{\Delta_{k}}}\ll1\ \ \ \ (j,k=1,2,...,N)\label{eq:dispcond}
\end{equation}
allows one to perform a Magnus expansion \cite{Magnus1954} on the
time-evolution operator $\hU^{I}(t)$ associated to $\hH^{I}(t)$,
which up to second order produces 
\begin{equation}
\hU^{I}(t)\approx{\rm e}^{-i\hH_{\text{eff}}^{I}(t)},\ \ \hH_{\text{eff}}^{I}(t) = \Lambda_{\!{_N}}t\hsgp\hsgm +\frac{\hsgz}{2}\hat{M}(t),\label{eq:Umagexp-1}
\end{equation}%
where $\Lambda_{\!{_N}} =  \sum_{j=1}^{N}{g_{j}^{2}}/{\Delta_{j}}$ 
is the resulting energy shift on the qubit, and the bosonic part of Eq.~(\ref{eq:Umagexp-1}) is given by
\begin{equation}\label{opM}
\hat{M}(t) = \sum_{j,k=1}^{N} m_{jk}(t) \hat a_j^\dag \hat a_k,  
\end{equation}
where 
\begin{equation} \label{eq:omegajk}
m_{jk}(t) = i\frac{g_{j}g_{k}}{\Delta_{j}\Delta_{k}}\left(\Delta_{j}+\Delta_{k}\right)
              \left(\frac{1-e^{i\left(\omega_{j}-\omega_{k}\right)t}}{\omega_{j}-\omega_{k}}\right).
\end{equation}
The complete derivation of Eq.~(\ref{eq:Umagexp-1}) and numerical comparisons with the dynamics governed by Hamiltonian~(\ref{eq:HJCN}) are shown in the Appendixes~\ref{App:EffDin} and~\ref{App:Num}, respectively. It follows from  Eq.~(\ref{opM}) that, apart from energy shifts, the  \textit{dispersive condition} stated in Eq.~(\ref{eq:dispcond}) 
also promotes interaction among the modes, see Fig.~\ref{fig:model}. 
Quite importantly, this interaction is dependent on the state of the qubit through $\hat{\sigma}_z$ in Hamiltonian~(\ref{eq:Umagexp-1}).  
Also, given that $[\frac{\omega_{0}}{2} \hat \sigma_z, \hH_{\text{eff}}^{I}(t) ] = 0$, there are no population changes in the eigenstates of $\hat{\sigma}_z$ so that the resulting dynamics will be energy conserving and only changes in the qubit coherence will be observed.

%

The initial state of the global system is assumed to be in the form
$\hrh(0)=\hrh_{\!_{S}}\otimes\hrh_{\!{_E}}$, where the subscript $S$ refers to the qubit and $E$ to the $N$-mode environment. We focus on the reduced dynamics
of the qubit described by $\hrh_{\!_{S}}(t)=\Tr_{E}\hrh(t)$. The time evolution of the coherence in the original Schrödinger picture is  
\begin{equation}
\rho_{01}(t):=\langle0|\hat{\rho}_{S}(t)|1\rangle=\rho_{01}(0) \, e^{-i\Lambda_{\!{_N}}t}\, r_{\!{_N}}(t),\label{eq:rho01t}
\end{equation}
with 
\begin{equation}
r_{\!{_N}}(t)=\Tr_{E}\left[\hrh_{\!{_E}} \, e^{-i\,\hM(t)}\right].\label{eq:rN1}
\end{equation}
This quantity completely characterizes the open dynamics of the qubit in our model, regardless of the number of modes. Furthermore, $\n{r_N(t)}$ is proportional to the well-known $l_1$-norm measure of coherence \cite{Baumgratz2013}. If the environment is initially at the zero-temperature vacuum state $\hrh_{\!{_E}}=\kb{0_{1},\dots,0_{N}}{0_{1},\dots,0_{N}}$, Eq.~(\ref{eq:rN1}) reveals that the dispersive condition inhibits correlations between the qubit and the modes so that $r_{\!{_N}}(t)=1$. 
This is valid regardless of the frequency distribution of the modes as long as the dispersive condition Eq.~(\ref{eq:dispcond}) is observed.
\section{Degenerate modes in thermal equilibrium}

Now we consider a thermal environment where $\hrh_{\!{_E}}$ is the product of $N$ Gibbs states, one for each mode, and equal temperature $T$ for all modes. In this case, the thermal occupation of mode $k$ is $\bar{n}_k=\left(e^{\omega_k/T}-1\right)^{-1}$, where we considered the Boltzmann constant $k_B=1$. We start the analysis of our model by first considering the degenerate case, where all modes have the same angular frequency $\omega_{j} = \omega$ $(j = 1,2,...,N)$ and therefore the same thermal occupation $\bar{n}_j = \bar{n}=\left(e^{\omega/T}-1\right)^{-1}$
$(j = 1,2,...,N)$. In this case, the coherence will depend only on the total shift $\Lambda_N$ 
and the thermal occupation number $\bar{n}$ through (see Appendix~\ref{App:Deg})
\begin{equation}
r_{\!{_N}}(t) = \frac{e^{i\Lambda_{\!{_N}}t}}{\cos\left(\Lambda_{\!{_N}}t\right)+i\left(2\bar{n}+1\right)\sin\left(\Lambda_{\!{_N}}t\right)}.\label{eq:rNdeg}
\end{equation}
In this case, $r_{\!{_N}}(t)$ is periodic in time, with the particular feature that its frequency of oscillation increases with the number of modes due to
$\Lambda_{\!{_N}} = (\omega_0 - \omega)^{-1} \sum_{j=1}^{N} g_j^2$. 
Consequently, the time taken for the total revival of the initial qubit state shortens as the number of degenerate modes increases. 
This behavior is different from what is observed in canonical models of decoherence, where recurrences are usually delayed with augmentation of the number of subsystems. It is important to remark that the periodicity of $r_N(t)$ here is not a consequence of the Magnus expansion but is, indeed, a natural feature arising from the Hamiltonian~(\ref{eq:HJCN}) provided that the dispersive condition~(\ref{eq:dispcond}) is fulfilled, as shown numerically in Appendix~\ref{App:Num}. The Magnus expansion here reveals only the effective structure of the system-environment coupling. Physically, the presence of degenerate modes in the environment is equivalent to the dispersive interaction between the qubit and a single mode with adjustable coupling constants. Such two-body interaction naturally produces periodic $r_N(t)$.

An insight about this feature can be obtained through the diagonalization of operator $\hM(t)$ defined
in Eq.~(\ref{opM}). Since $\hM(t)$ is Hermitian, there might be a time-dependent
unitary operator $\hV(t)$ which diagonalizes it. Given that Hamiltonian~(\ref{opM}) does not promote squeezing, i.e., 
there are no terms in the form $\ha_j \ha_k$ or $\hdgg{a}_j \hdgg{a}_k$, we can write
\begin{equation}
\hM_{d}(t) =\hV(t)\hM(t)\hdgg V(t) =\sum_{j=1}^{N}\epsilon_{j}(t)\hdgg a_{j}\ha_{j}. \label{eq:Md}
\end{equation}
%
In what follows, $\boldsymbol{M}(t)$ is the $N\times N$ Hermitian matrix  
whose entries are $m_{jk}(t)$ given by Eq.~(\ref{eq:omegajk}). Also, $\boldsymbol{V}(t)$ is a unitary matrix such that 
$\boldsymbol{V}(t)^\dagger \boldsymbol{M}(t) \boldsymbol{V}(t) = {\rm Diag}[\epsilon_{1}(t),...,\epsilon_{N}(t)]$. Then,
\begin{equation} \label{eq:diagRel}
\hV(t)\ha_{j}\hdgg V(t) = \sum_{k =1}^N \boldsymbol{V}_{\!\!jk}(t) \ha_k \,\,\,\,\, (j = 1,...,N). 
\end{equation}
For the degenerate case, $m_{jk}(t)=2g_jg_k t/(\omega_0 - \omega)$, and consequently $\boldsymbol{M}(t)$ is actually a dyadic such that its single non-null eigenvalue is given by $\epsilon_1(t)=\Lambda_N t$. Therefore, the qubit is only effectively coupled to a single mode, and the remaining degenerate modes only contribute to the intensity of such interaction. This constitutes an effective way of controlling the single-mode dispersive regime within the scope of the rotating-wave approximation, since the values of $g_j$ are kept fixed. Naturally, the above development is only valid for finite $N$. 
\vspace*{-3mm}
\section{Nondegenerate modes in thermal equilibrium}

The behavior of the coherence is enriched when nondegenerate modes are used due to the oscillating behavior of $m_{jk}(t)$. We now detail the derivation of analytical results for the coherence, which includes this case.

The evaluation of Eq.~(\ref{eq:rN1}) is facilitated by evaluating the trace in the coherent states basis and by expressing $\hrh_{\!{_E}}$ in the
$P$ representation \cite{Glauber1963,Sudarshan1963}, i.e.,
$\hrh_{\!{_E}}=\int \! d^{2N}\! \boldsymbol{\alpha} \, P(\boldsymbol{\alpha})\kb{\boldsymbol{\alpha}}{\boldsymbol{\alpha}}$,
where $\ket{\boldsymbol{\alpha}}=\ket{\alpha_{1},\dots,\alpha_{N}}$
is a $N$-mode coherent state. Through the diagonalization of $\hM(t)$, one finds
\begin{equation}
r_{\!{_N}}(t)=\int\! d^{2N}\!\boldsymbol{\alpha} \, P(\boldsymbol{\alpha})\mel{\boldsymbol{\alpha}}{\hdgg V(t)e^{-i\hM_{d}(t)}\hV(t)}{\boldsymbol{\alpha}},\label{eq:rN2}
\end{equation}
and further analytical progress depends now on how we deal with
the action of $\hat{V}(t)$ on $\ket{\boldsymbol{\alpha}}$.
Notice that Eq.~(\ref{eq:diagRel}) implies that
$\hat{V}(t)\ket{\boldsymbol{\alpha}}$ is an eigenvector of $\hat a_j$ with eigenvalue ${\alpha}'_{j}(t)$ given by
\begin{equation}
{\alpha}'_{j}(t)=\sum_{k=1}^{N}{\boldsymbol V}_{\!\!kj}^\ast (t)\, \alpha_{k}.\label{eq:alphal}
\end{equation}
This means that 
$\hV(t)\ket{\boldsymbol{\alpha}}=\ket{\boldsymbol{\alpha}'(t)}=\ket{{\alpha}'_{1}(t),\dots,{\alpha}'_{N}(t)}$ is itself a factorized multimode coherent state.  
Using now
$e^{-i\epsilon_{j}(t)\ha_{j}^{\dagger}\ha_{j}}\ket{\alpha_{j}'(t)}=\ket{\alpha_{j}'(t)e^{-i\epsilon_{j}(t)}}$ and the thermal distribution
\begin{equation}\label{pf}
P(\boldsymbol{\alpha})=\frac{e^{-\sum_{j=1}^{N}\n{\alpha_{j}}^{2}/\bar{n}_{j}}}{\pi^{N}\prod_{j=1}^{N}\bar{n}_{j}},
\end{equation}
we end up with  
\begin{equation}
r_{\!{_N}}(t)=\frac{1}{\prod_{j=1}^{N}\bar{n}_{j}}\frac{1}{\sqrt{\det\boldsymbol{A}(t)}},\label{eq:rN3}
\end{equation}
where $\boldsymbol{A}(t)$ is the $2N\times2N$ complex matrix 
\begin{equation}
\boldsymbol{A}(t)=\left(\begin{array}{cccc}
y_{1}(t)\boldsymbol{I} & \boldsymbol{W}_{\!\!12}(t) & \dots & \boldsymbol{W}_{\!\!1N}(t)\\
\boldsymbol{W}_{\!\!12}^{\top}(t) & y_{2}(t)\boldsymbol{I} & \dots & \boldsymbol{W}_{\!\!2N}(t)\\
\vdots & \dots & \ddots & \vdots\\
\boldsymbol{W}_{\!\!1N}^{\top}(t) & \boldsymbol{W}_{\!\!2N}^{\top}(t) & \dots & y_{N}(t)\boldsymbol{I}
\end{array}\right). \label{eq:A}
\end{equation}
The compact and analytical form of the function $r_N(t)$ in Eq.~(\ref{eq:rN3}) results from the evaluation of a $2N$-dimensional Gaussian integral. Still, in Eq.~(\ref{eq:A}), $\boldsymbol{I}$ denotes the $2\times2$ identity matrix and
\begin{equation}
\boldsymbol{W}_{\!\!jk}(t)=\left(\begin{array}{cc}
-u_{jk}(t) & -v_{jk}(t)\\
v_{jk}(t) & -u_{jk}(t)
\end{array}\right), \label{eq:Wjk}
\end{equation}
with
\begin{align}\label{eq:auxfunc}
y_{j}(t) & =\bar{n}_{j}^{-1}\left(\bar{n}_{j}+1\right)-\sum_{l=1}^{N}e^{-i\epsilon_{l}(t)}\ns{\boldsymbol{V}_{\!\!jl}(t)},\nonumber\\
u_{jk}(t) & =\sum_{l=1}^{N}e^{-i\epsilon_{l}(t)}\text{\ensuremath{\text{Re}}}\left[\boldsymbol{V}_{\!\!jl}^{*}(t)\boldsymbol{V}_{\!\!kl}(t)\right],\\
v_{jk}(t) & =\sum_{l=1}^{N}e^{-i\epsilon_{l}(t)}\text{Im}\left[\boldsymbol{V}_{\!\!jl}^{*}(t)\boldsymbol{V}_{\!\!kl}(t)\right].\nonumber
\end{align}
Note that we are not bound to the case of identical thermal occupations for each mode since $\bar{n}_j$  is a free parameter in~(\ref{pf}). 

\textit{Examples}. We start our analysis with the first nontrivial case which is $N=2$. For this choice, Eqs.~(\ref{eq:rN3})--(\ref{eq:auxfunc}) allow one to obtain
\begin{align}\label{eq:r2}
r_{2}(t)&=e^{i\epsilon_{+}(t)}\left[\bar{n}_{1}\bar{n}_{2}e^{-i\epsilon_{+}(t)}+\left(\bar{n}_{1}+1\right)\left(\bar{n}_{2}+1\right)e^{i\epsilon_{+}(t)}\right.\nonumber\\ 
{}&\;\;\;\;\left.-\left(\bar{n}_{1}\bar{n}_{2}+\bar{n}_{1}\ns{\boldsymbol{V}\!\!_{11}(t)}+\bar{n}_{2}\ns{\boldsymbol{V}\!\!_{12}(t)}\right)e^{-i\epsilon_{-}(t)}\right.\nonumber\\
{}&\;\;\;\;\left.-\left(\bar{n}_{1}\bar{n}_{2}+\bar{n}_{1}\ns{\boldsymbol{V}\!\!_{12}(t)}+\bar{n}_{2}\ns{\boldsymbol{V}\!\!_{11}(t)}\right)e^{i\epsilon_{-}(t)}\right]^{-1},
\end{align}
where $\epsilon_{\pm}(t)=\left[\epsilon_{1}(t)\pm\epsilon_{2}(t)\right]/2$, and $\epsilon_{1,2}(t)$ are the eigenvalues of $\hat{M}(t)$. Moreover, 
following the described protocol, 
the entries of $\boldsymbol{V}(t)$ satisfy (omitting the time dependence) $\ns{\boldsymbol{V}\!\!_{11}}=\ns{\boldsymbol{V}\!\!_{22}}=\ns{m_{12}}\bigl[\left(\epsilon_{1}-m_{11}\right)^{2}+\ns{m_{12}}\bigr]^{-1}$ and $\ns{\boldsymbol{V}\!\!_{12}}=\ns{\boldsymbol{V}\!\!_{21}}=\left(\epsilon_{1}-m_{11}\right)^{2}\bigl[\left(\epsilon_{1}-m_{11}\right)^{2}+\ns{m_{12}}\bigr]^{-1}$. One can easily check that Eq.~(\ref{eq:r2}) reduces to Eq.~(\ref{eq:rNdeg}) when $\bar{n}_1=\bar{n}_2$ and $\omega_1=\omega_2=\omega$. For nondegenerate modes, one finds $\epsilon_{+}(t)=\Lambda_2 t$ and $\epsilon_{-}(t)=\tfrac{g_{1}g_{2}}{\Delta_{1}\Delta_{2}}t\bigl[(\tfrac{g_{1}}{g_{2}}\Delta_{2}-\tfrac{g_{2}}{g_{1}}\Delta_{1})^2 +(\Delta_{1}+\Delta_{2})^{2}f^2(t)\bigr]^{1/2}$, with $f(t)=\sin\left[\left(\omega_{1}-\omega_{2}\right)t/2\right]\left[\left(\omega_{1}-\omega_{2}\right)t/2\right]^{-1}$. Therefore, the time-dependence on the induced mode-mode coupling tends to disturb the periodicity of $r_2(t)$ since $\epsilon_{\pm}(t)/t$ are in general dynamically incommensurate. This confers on the coherence of $\hrh_S(t)$ an interesting and nontrivial dynamics which can be engineered through the choices of the system parameters.

\begin{figure*}
\includegraphics[width=0.95\textwidth]{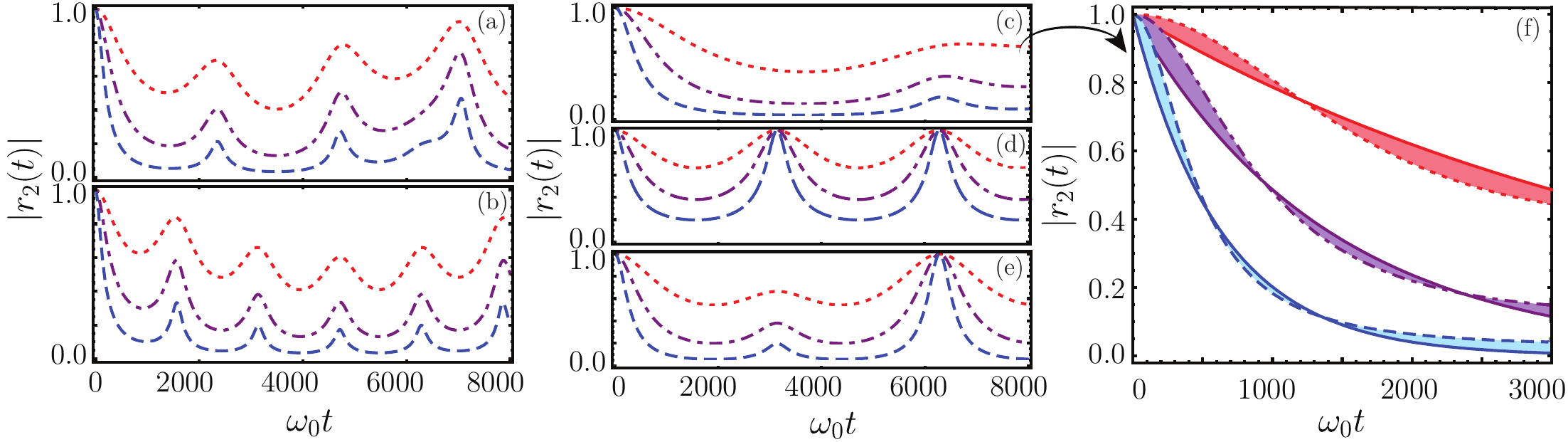} 
\subfloat{\label{fig:r2a}}
\subfloat{\label{fig:r2b}}
\subfloat{\label{fig:r2c}}
\subfloat{\label{fig:r2d}}
\subfloat{\label{fig:r2e}}
\subfloat{\label{fig:r2f}}
\vspace*{-3mm}
\caption{(Color online) Qubit coherence indicator $\n{r_2(t)}$ as a function of the dimensionless time $\omega_0 t$. Red dotted curves: $T=0.5 \omega_0$; purple dot-dashed curves: $T=1.0\omega_0$; blue dashed curves: $T=2.0\omega_0$. In the left panels, the modes have $\omega_1=0.8\omega_0$ and $\omega_2=0.7\omega_0$, whereas the coupling constants are (a) $g_{1}=0.01\omega_0$ and $g_{2}=0.02\omega_0$; and (b) $g_{1}=0.02\omega_0$ and $g_{2}=0.01\omega_0$. In the center panels, coupling constants are fixed at $g_1=g_2=0.01 \omega_0$, whereas for modes (c) $\omega_1=0.8\omega_0$ and $\omega_2=0.7\omega_0$; (d) $\omega_1=\omega_2=0.8\omega_0$ [degenerate case that follows from Eq.~(\ref{eq:rNdeg})]; and (e) $\omega_1=0.8\omega_0$ and $\omega_2=0.9\omega_0$. The right panel (f) shows the short-time behavior of $\n{r_2(t)}$ for the same parameters as in (c). In this case, the colored areas indicate the deviation of $\n{r_2(t)}$ from exponential decays $e^{-2\gamma t}$ with rates $\gamma=1.2\times10^{-4} \omega_0$ (red), $\gamma=3.6\times10^{-4} \omega_0$ (purple), and $\gamma=8.0\times10^{-4} \omega_0$ (blue).\label{fig:r2}}
\end{figure*}

Figure~\ref{fig:r2} shows $\n{r_2(t)}$ as a function of the dimensionless time $\omega_0 t$ for different choices of frequencies $\omega_j$, coupling constants $g_j$, and temperature $T$. As indicated in the left panels, keeping the modes slightly detuned from each other is sufficient to make $\n{r_2(t)}$ exhibit irregular oscillations due to $\epsilon_{+}(t)\neq\epsilon_{-}(t)$ and $\bar{n}_1\neq \bar{n}_2$ in Eq.~(\ref{eq:r2}). In the center panels, for fixed coupling constants $g_1$ and $g_2$, and also for fixed $\omega_1$, we emphasize the sensibility of $\n{r_2(t)}$ with the variation of $\omega_2$. In all plots, the parameters $g_j/\Delta_j$ essentially dictate the speed of the oscillations, which is somehow related to the intensity of the system-environment coupling through Eqs.~(\ref{eq:Umagexp-1}--\ref{eq:omegajk}). On the other hand, the thermal occupations $\bar{n}_j$ determine the amplitudes. Physically, larger values of $\bar{n}_j$ cause the qubit dynamics to be more subjected to the high excited levels of the modes. Consequently, the effects of decoherence are more pronounced and the qubit tends to reach states with a lower degree of purity.

Besides its suitability to the study of decoherence, the function $\n{r_{\!{_N}}(t)}$ also reveals non-Markovianity in the dynamics through its attempts to recur. For a pure dephasing model, such as the one considered here, there is a one-to-one correspondence between $\n{r_{\!{_N}}(t)}$ and the distinguishability of the pair of initial states consisting of eigenstates of $\hat{\sigma}_x$. Any dynamical increment of distinguishability between these two initial states indicates non-Markovianity as it is a consequence of information backflow from the environment to the system \cite{Breuer2009,Wismann2012,Breuer2016}. All plots in the left and center panels of Fig.~\ref{fig:r2} signalize non-Markovianity, which is a direct consequence of the finiteness of the environment \cite{Schlosshauer2008}. 

For engineering of qubit dephasing, with fixed $g_j$, the nondegenerate case [see Figs.~\subref*{fig:r2c} and~\subref*{fig:r2e}] is more appropriate than the degenerate case [see Fig.~\subref*{fig:r2d}], since it allows one to achieve lower values of $\n{r_2(t)}$ and to maintain them for longer times, which promotes a delay of the complete recurrence.
The short-time regime is particularly interesting as it can be used to reproduce qualitatively the influence of canonical types of environment on the qubit dynamics, from which a monotonic decay of the coherence is expected. In our setup, even for very small environments ($N=2$), one can see that proper adjustments of $g_j$, $\Delta_j$, and $T$ may lead to an approximate monotonic decay of the coherences, typical, for instance, of the weak interaction of the qubit with a macroscopic pure dephasing bath [Fig.~\subref*{fig:r2f}]. In particular, the temperature $T$ radically affects the quality of the emulation, as mentioned previously. It is important to remark that the dynamical map as a whole is non-Markovian and there will be attempts of recurrence for a finite time. 

\begin{figure} 
\subfloat{\label{fig:tmaxa}} 
\subfloat{\label{fig:tmaxb}}
\includegraphics[width=\linewidth]{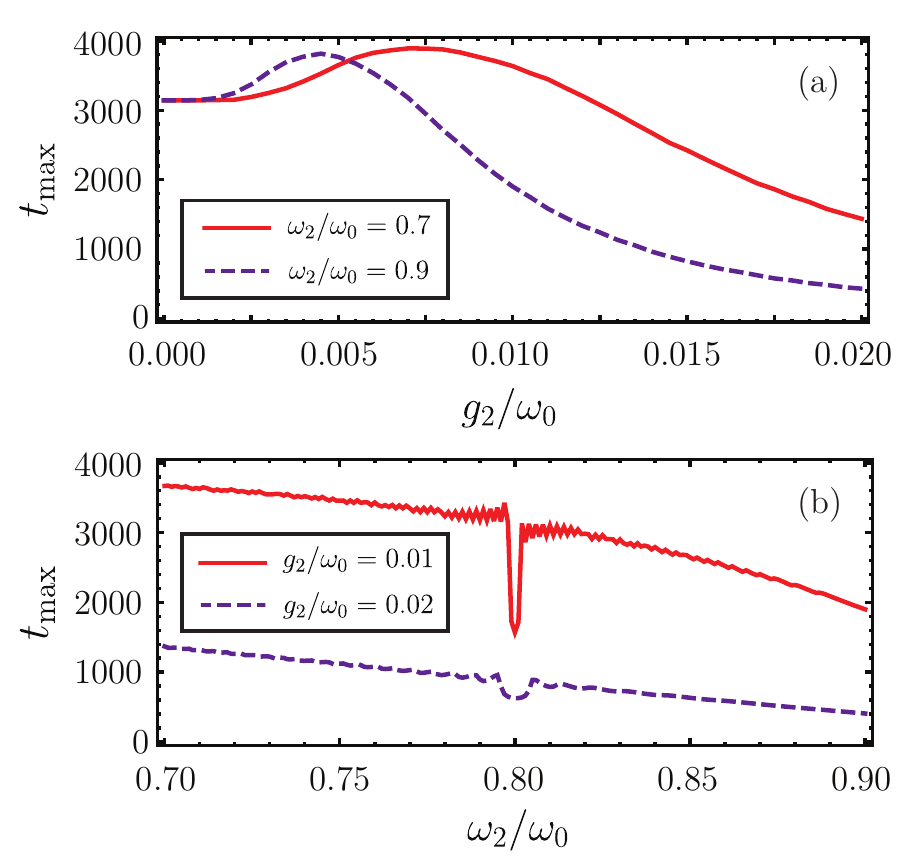}
\vspace*{-3mm} 
\caption{(Color online) Plots of $t_{\text{max}}$ (in units of $\omega_0^{-1}$) for $N=2$. We choose $g_1=0.01\omega_0$, $\omega_1=0.8\omega_0$, and $T=0.5\omega_0$. The sensitivity of $t_{\text{max}}$ with the variations of $g_2$ and $\omega_2$ are shown in panels (a) and (b), respectively.
\label{fig:tmax}} 
\end{figure}

Further investigation on the behavior of $\n{r_{\!{_N}}(t)}$ also allows one to extract its maximal time of monotonic decay, $t_{\text{max}}$. This corresponds to the instant of time before the first recurrence of the function $\n{r_{\!{_N}}(t)}$, or equivalently, its first local minimum. For the degenerate case, such times can be obtained analytically from Eq.~(\ref{eq:rNdeg}), and it is given by $t_{\text{max}}=\pi/(2\Lambda_N)$. Considering the nondegenerate case and still with $N=2$, in Fig.~\ref{fig:tmax} we numerically show the dependence of $t_{\text{max}}$ on the variation of $g_2$ and $\omega_2$ for fixed values of $g_1$, $\omega_1$, and $T$. From the panel~\subref*{fig:tmaxa}, it is possible to see that $t_{\text{max}}$ is kept almost constant for very small values of $g_2$, as this situation resembles the coupling to a single mode. For intermediate values of $g_2$ but still smaller than $g_1$, $t_\text{max}$ can be delayed for more than $20\%$ of its values obtained with very small $g_2$. Then, as $g_2$ approaches $g_1$ and eventually becomes larger, 
$t_{\text{max}}$ decreases as a result of the faster oscillations of $\n{r_2(t)}$ generated by the higher values of $g_2/\Delta_2$. The effects of such increments to the behavior of $t_{\text{max}}$ are also present in panel~\subref*{fig:tmaxb}, where one notices a smooth decay of $t_{\text{max}}$ for the chosen range of $\omega_2$. The exception occurs in the region close to the resonance ($\omega_2\approx\omega_1=0.8\omega_0$). In this case, $t_{\text{max}}$ falls abruptly, evidencing that nonresonant modes tend to promote prolonged monotonic decays of $\n{r_2(t)}$ once one has fixed the coupling constants $g_j$. Mathematically, this sudden change is a consequence of the rich behavior of $r_{N}(t)$ even for $N=2$. Just before the resonance, the curve $\n{r_{2}(t)}$ develops an
inflection at a point $t < t_{\rm max}$, which becomes a new minimum of $\n{r_{2}(t)}$ as long as $\omega_2$
approaches $\omega_1$.

\begin{figure*} 
\subfloat{\label{fig:rNa}} 
\subfloat{\label{fig:rNb}}
\subfloat{\label{fig:r3c}}
\subfloat{\label{fig:r4d}}
\subfloat{\label{fig:r5e}}
\subfloat{\label{fig:r10f}}
\includegraphics[width=\textwidth]{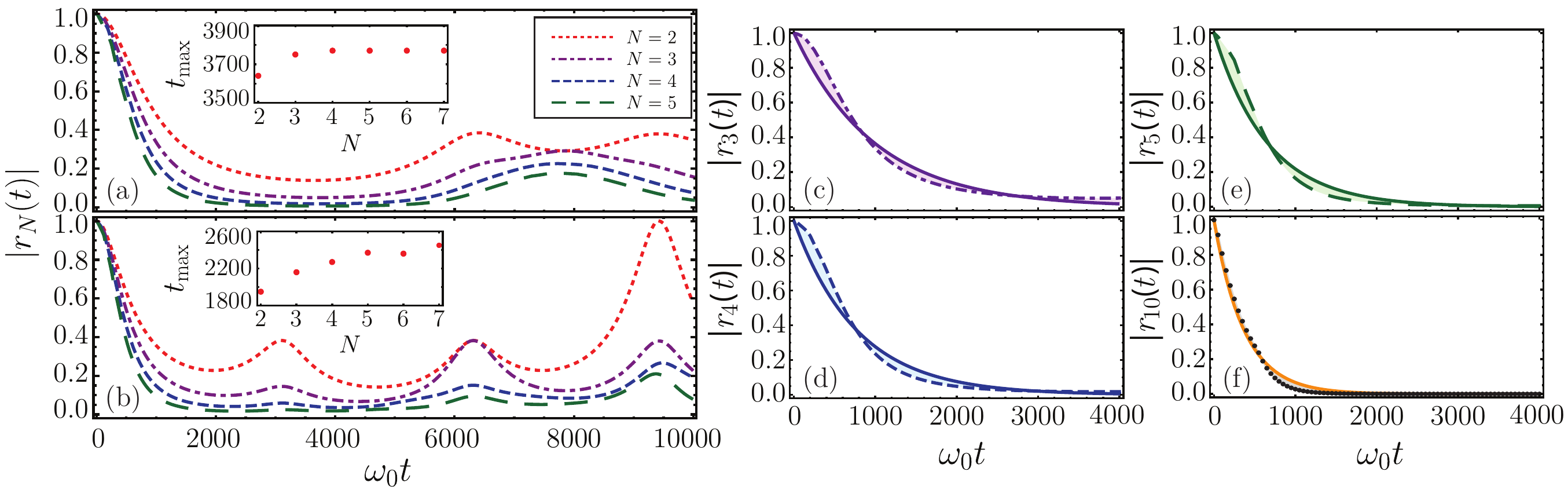}
\vspace*{-3mm} 
\caption{(Color online) Plots of $\n{r_{\!{_N}}(t)}$ as a function of the dimensionless time $\omega_0 t$ for different values of $N$. Temperature and coupling constants are given respectively by $T=1.0\omega_0$ and $g_j=0.01\omega_0$. Frequencies of the modes are equally spaced in the interval $[\omega_1,\omega_N]$, with $\omega_1=0.7\omega_0$, $\omega_N=0.8\omega_0$ [panel (a)], and $\omega_N=0.9\omega_0$ [panel (b)]. The inset plots of panels (a) and (b) show $t_{\text{max}}$ (in units of $\omega_0^{-1}$) as a function of $N$. Panels (c)--(f) show the short-time behavior of $\n{r_{\!{_N}}(t)}$ for parameters chosen as in panel (a). Points, dashed, and dot-dashed lines represent $\n{r_{\!{_N}}(t)}$ calculated through Eq.~(\ref{eq:rN3}), whereas solid lines are exponential fits $e^{-2\gamma t}$, with (c) $\gamma=5.1\times10^{-4}\omega_0$, (d) $\gamma=6.4\times10^{-4}\omega_0$, (e) $\gamma=7.6\times10^{-4}\omega_0$, and (f) $\gamma=1.4\times10^{-3}\omega_0$.
\label{fig:rN}} 
\end{figure*}

We examine now the behavior of $\n{r_{\!{_N}}(t)}$ for larger environments. According to Eqs.~(\ref{eq:rN3})--(\ref{eq:auxfunc}), the inclusion of more modes in the environment results in additional oscillating terms in $r_{\!{_N}}(t)$ with amplitudes consisting of products of $\bar{n}_j$. If such modes are nondegenerate, more steep depletions of $\n{r_{\!{_N}}(t)}$ are expected when $N$ increases, since all eigenvalues of the matrix $\boldsymbol{M}(t)$, $\epsilon_j(t)$, are non-null in general. Such features can be seen in Fig.~\ref{fig:rN}, where $g_j$ and $T$ are kept fixed and the frequencies of the modes are equally spaced in the interval $[\omega_1,\omega_N]$. From the inset plots of Figs.~\subref*{fig:rNa} and~\subref*{fig:rNb}, it is also possible to notice that $t_{\text{max}}$ is barely affected by the augmentation of modes, and such time is delayed when greater detunings $\Delta_j$ are present [Fig.~\subref*{fig:rNa}]. In addition, the presence of more (distinguished) frequencies in the environment attenuates the revivals of $\n{r_{\!{_N}}(t)}$, stabilizing the coherence at low values for longer times. Consequently, $\n{r_{\!{_N}}(t)}$ takes longer to recur completely to its initial value, which indicates that, for the chosen parameters, information backflow to the system is prevented as $N$ increases, in contrast to what is observed in the complete degenerate case [see Eq.~(\ref{eq:rNdeg}) and Fig.~\subref*{fig:r2d}]. Indeed, the STMD of the qubit coherence is prone to resemble an exponential decay as the nondegenerate environment becomes sufficiently large [see Figs.~\subref*{fig:r3c}--\subref*{fig:r10f}].

Now for some comments on the feasibility of observation of our results in a real setup. In circuit QED, temperatures about a few dozens or hundreds of mK are easily achieved using dilution refrigerators \cite{Uhlig2015}. Such temperatures are consistent with our choice $T\approx\omega_0$, i.e., a thermal mode energy comparable to the typical qubit energies $\omega_0/ 2\pi \approx 10$ GHz \cite{Blais2007}. This is the temperature regime adopted in Figs.~\ref{fig:r2} and~\ref{fig:rN}. Also, in these architectures, qubit coherence times have reached a few microseconds \cite{Paik2011}. Again, given that $\omega_0/2\pi\approx 10$ GHz, a time window of $3000/\omega_0$ would correspond to some fractions of microseconds, which is well inside the coherence time of the qubit. 

\section{Conclusion}

We have employed the multimode dispersive Jaynes-Cummings interaction to induce an energy-preserving open dynamics on the qubit. Closed-form expressions for the qubit coherence valid for $N$ environmental modes in thermal equilibrium were obtained. Our investigation is fully within the grasp of current quantum technologies, with particular interest for circuit QED implementations. Here, the modes are assumed to be single-mode resonators such that we only considered finite $N$. There is an ongoing discussion about potential divergencies in a multimode resonator as $N\rightarrow\infty$ \cite{Gely2017,Malekakhlagh2017}. We also discussed issues such as the possibility of production of short-time monotonic decay of the coherences according with the parameters of the model, which might be useful for the simulation of qubit dephasing and the investigation of its role in quantum protocols and operations. A future extension of our work might consider the analysis of the long-time behavior of the coherences when natural dephasing and dissipation take place. Such studies are relevant for the characterization of steady-state properties of the system \cite{Guarnieri2018,Wang2018,Wang2018a,Nicacio2016,Barreiro2011}. Another possibility may focus on the ultrastrong-coupling regime where $g_j/\omega_j>1$ \cite{Bosman2017}. In this case, a generalized dispersive regime beyond the rotating-wave approximation will have to be considered. Finally, one can also think of fermionic systems where the transition from finite to infinite elements in the environment have been recently investigated \cite{Thingna2017}. 

\section*{Acknowledgements} 

W.S.T. acknowledges Fundação de Amparo à Pesquisa
do Estado de São Paulo (FAPESP) for financial support
through Grant No. 2017/09058-2. F.N. is partially supported
by the Brazilian agency CAPES through the PNPD program.
F.L.S. acknowledges partial support from CNPq (Grant No.
302900/2017-9). F.N. and F.L.S. also acknowledge partial
support from the Brazilian National Institute of Science and
Technology of Quantum Information (INCT-IQ).

\setcounter{equation}{0}
\setcounter{table}{0}
\makeatletter
\renewcommand{\theequation}{A\arabic{equation}}

\appendix
\section{Derivation of the effective dynamics}\label{App:EffDin}

Here we detail the derivation of Eq.~(\ref{eq:Umagexp-1}) from the main text, 
which describes the effective evolution of a qubit interacting dispersively 
with $N$ electromagnetic modes. 

The time-evolution operator for the interaction-picture Hamiltonian (\ref{eq:HJCI}) 
is expressed as a Magnus series \cite{Magnus1954,Tannor2007} in terms of the anti-Hermitian operator 
$\hOm(t)$, {i.e.}, 
%
\begin{equation}
\hU^{I}(t)=\exp\left[\hOm(t)\right],\;\;\;\;\hOm(t)=\sum_{n=1}^{\infty}\hOm_{n}(t). \label{eq:UintpicMag}
\end{equation}
Conveniently, the Magnus expansion perturbatively produces a unitary operator for any desired order, 
which is not true for the usual Dyson series \cite{Tannor2007}. 
Following \cite{Magnus1954,Tannor2007} and using the Hamiltonian (\ref{eq:HJCI}), the first two terms of this expansion are
\begin{eqnarray}
\!\!\!\!\hOm_{1}(t) & = &-i\int_{0}^{t}dt_1\hH^{I}(t_1) \nonumber \\
            & = & \sum_{j=1}^{N}\frac{g_{j}}{\Delta_{j}}\left[\left(1-e^{i\Delta_{j}t}\right)\hsgp\ha_{j}-\text{H.C.}\right], \label{eq:Omn1} \\ 
\!\!\!\!\hOm_{2}(t) & = &  -\frac{1}{2}\int_{0}^{t}dt_1\int_{0}^{t_1}dt_2\left[\hH^{I}(t_1),\hH^{I}(t_2)\right] \nonumber \\
            & = & -i\Lambda_{\!{_N}} t \,\hsgp\hsgm -i\frac{\hsgz}{2}\hM(t) + \mathcal O^2\!\left(\tfrac{g_j}{\Delta_j}\right), \label{eq:Omnt}         
\end{eqnarray}
where $\Lambda_{\!{_N}}$ and $\hM(t)$ are both defined in the main text, respectively  
below Eq.~(\ref{eq:Umagexp-1}) and in Eq.~(\ref{opM}). 

The generalized dispersive condition $\n{g_j/\Delta_k}\ll1$ in (\ref{eq:dispcond}) allows one 
to neglect the higher-order terms in (\ref{eq:Omnt}). The same applies to the terms $\hOm_n(t)$ with $n>2$ in (\ref{eq:UintpicMag}), 
since they involve higher-order commutators \cite{Magnus1954,Tannor2007} which give rise to products of $g_j/\Delta_k$. 
Therefore, up to second order, the time-evolution operator in (\ref{eq:UintpicMag}) for the interaction-picture Hamiltonian (\ref{eq:HJCI}) is well approximated by $\hU^{I}(t)\approx e^{\hOm_{1}(t)+\hOm_{2}(t)}$.
Now, another important assumption is to neglect the influence of $\hOm_{1}(t)$ on the dynamics, and this can be understood as follows. Considering an initial state in the interaction-picture $\ket{\psi}$, the suppression of $\hOm_{1}(t)$ on the dynamics would demand that $\bra{\psi}e^{\hOm_{1}(t)+\hOm_{2}(t)}\ket{\psi}\approx\bra{\psi}e^{\hOm_{2}(t)}\ket{\psi}$. Roughly speaking, this is obtained if $\n{\bra{\psi}\hOm_{1}(t)\ket{\psi}}$ is sufficiently small, which is assured in our case, since $\hOm_{1}(t)$ is an oscillatory function of the time with amplitudes $\n{g_j/\Delta_j}\ll 1$, 
see Eq.~(\ref{eq:Omn1}). To some extent, dropping out such terms is what witnesses the dispersive feature of the model:
it inhibits energy exchange between the qubit and the modes due to the elimination of terms with 
$\hsgp\ha_{j}$ and $\hsgm\hdgg{a}_j$ in (\ref{eq:Omn1}). The same is found using other perturbative approaches for the study of the dispersive regime, e.g., \cite{Zhu2013}. 

On the other hand, the main contribution to the effective description comes from the 
energy exchange among the modes; these are represented by the crossing terms 
$\hdgg{a}_j\ha_k$ in $\hOm_{2}(t)$ throughout the operator $\hat M(t)$ defined in (\ref{opM}). 
Such terms become essentially linear in time, provided that 
the mode detunings $\n{\omega_j-\omega_k}$ are not large, see Eq.~(\ref{eq:omegajk}). 
This last condition is also indirectly required for both the dispersive limit explained above and the rotating-wave approximation producing (\ref{eq:HJCN}) to hold. 

When these conditions are fulfilled, it is possible  to write an effective evolution for 
the dispersive limit as in (\ref{eq:Umagexp-1}). To certify the validity of the effective description for the parameters adopted in the main text, we now provide a numerical evaluation with the exact Hamiltonian and contrast it with the results using the effective dispersive one. 
\begin{figure*}[t]
\includegraphics[width=0.95\textwidth]{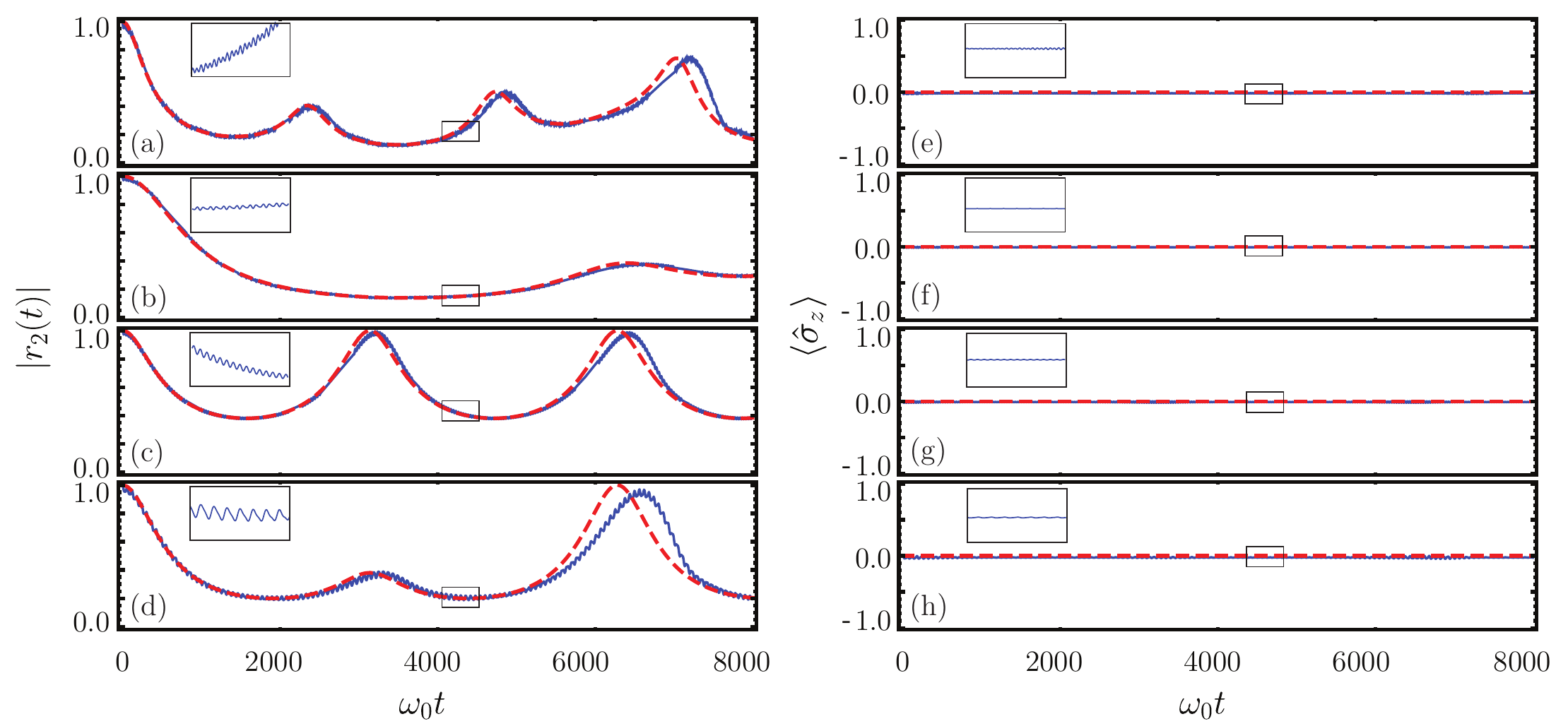} 
\subfloat{\label{fig:p3a}} 
\subfloat{\label{fig:p3b}} 
\subfloat{\label{fig:p3c}}
\subfloat{\label{fig:p3d}}
\subfloat{\label{fig:p3e}}
\subfloat{\label{fig:p3f}}
\subfloat{\label{fig:p3g}}
\subfloat{\label{fig:p3h}}
\vspace*{-3mm}
\caption{(Color online) Comparison of $\n{r_2(t)}$ on the left and $\ev{\hsgz}$ on the right obtained analytically 
through the Magnus expansion (red-dashed lines) with the corresponding curves obtained numerically from the exact 
Jaynes-Cummings Hamiltonian (blue-solid lines). 
Temperature is set at $T=1.0\omega_0$. Panels (a) and (e): $g_1=0.01\omega_0$, $g_2=0.02\omega_0$, $\omega_1=0.8\omega_0$, and $\omega_2=0.7\omega_0$. Panels (b) and (f): $g_1=g_2=0.01\omega_0$, $\omega_1=0.8\omega_0$, and $\omega_2=0.7\omega_0$. Panels (c) and (g): $g_1=g_2=0.01\omega_0$, $\omega_1=\omega_2=0.8\omega_0$ (degenerate case). Panels (d) and (h): $g_1=g_2=0.01\omega_0$, $\omega_1=0.8\omega_0$, and $\omega_2=0.9\omega_0$.  
The inset in each plot highlights the tiny oscillations in the exact dynamics.  
The initial state of the qubit is the eigenstate of $\hsgx$ with eigenvalue $1$.
\label{fig:p3}} 
\end{figure*}

\renewcommand{\theequation}{B\arabic{equation}}

{\section{Numerical checking of the effective dynamics}\label{App:Num}}

In order to check the validity of the approximations, in writing the effective evolution 
for the qubit in the dispersive limit, one can numerically determine the evolution governed by 
the Hamiltonian (\ref{eq:HJCN}). 
The numerical procedure consists in writing the matrix elements of that Hamiltonian with respect to  
the Fock basis of the modes Hilbert space. Since this is an infinite dimensional space, we perform 
a truncation on the space dimension conveniently choosing the number of Fock states.  
Specifically, {for the case of $N=2$ modes, we calculate $\n{r_2(t)}$ from Eq.~(\ref{eq:rN1}) and the mean value of $\ev{\hsgz}$, which are  respectively related to the $l_1$-norm coherence measure and population inversion for the qubit in the $\hsgz$-- eigenbasis.} 
The results are shown in Fig.~\ref{fig:p3}.  
%
%
%
For the chosen parameters, they are the same as in the main text, and the function $\n{r_2(t)}$ 
from the Magnus expansion is in close agreement with the numerical results; small deviations begin to be noticed only for long times ($\omega_0 t \sim 6000$), and this is more apparent in the cases with higher values of $\n{g_j/\Delta_k}$, as in Figs.~\subref*{fig:p3a}, \subref*{fig:p3c}, and \subref*{fig:p3d}. As expected, the validity of the effective model is progressively degraded as $\n{g_j/\Delta_k}$ becomes larger, and this is why the tiny oscillations in the numerical curves are also more pronounced in the mentioned plots. Furthermore, the numerical calculations show that values of $\ev{\hsgz}$ are practically kept constant, 
{indicating that qubit populations are preserved}. Indeed, such behavior is predicted by 
the dispersive model, as explained in the main text. 
\\

 

\renewcommand{\theequation}{C\arabic{equation}}

{\section{Calculation of \texorpdfstring{$r_{\!{_N}}(t)$}{rN(t)} for degenerate modes \label{App:Deg}}

In this section we detail the derivation of Eq.~(\ref{eq:rNdeg}) presented in the main text. 
This equation exhibits the function $r_{\!{_N}}(t)$ in (\ref{eq:rN1}) 
for a set of $N$ electromagnetic modes with the same frequency and in thermal equilibrium. 
The procedure is based on the Weyl-Wigner formalism and related to the tools developed in 
Refs.~\cite{Nicacio2017,Gosson2018} for the obtainment of the total phase acquired by a Gaussian state.

In our context, the frequency degeneracy causes $\hM(t)$ in (\ref{opM}) to be linear in time, since Eq.~(\ref{eq:omegajk}) becomes $m_{jk}(t) = 2g_jg_k t/(\omega_0 - \omega)$, and  $r_{\!{_N}}(t)$ is the average value of the metaplectic operator 
\begin{equation}
\hR(t)=e^{i\Lambda_{\!{_N}}t}e^{-\frac{it}{2} \hx^{\top} \boldsymbol{H}\hx }, \label{eq:met}
\end{equation}
{i.e.}, the unitary operator generated by the quadratic Hamiltonian $\frac{1}{2} \hx^{\top} \boldsymbol{H}\hx$. For convenience we have defined the $2N$-dimensional column vector $\hx=\bigl(\hq_1,\dots,\hq_N,\hp_1,\dots,\hp_N\bigr)^{\top}$, which is composed by the quadrature operators given by $\hq_j = \bigl(\hdgg{a}_j+\ha_j\bigr)/\sqrt{2}$ and 
$\hp_j=i\bigl(\hdgg{a}_j-\ha_j\bigr)/\sqrt{2}$. 
Also, $\boldsymbol{H}$ is the $2N\times 2N$ symmetric real matrix given by 
\begin{equation}
\boldsymbol{H} = \boldsymbol{G}\oplus\boldsymbol{G},  \,\,\, 
\boldsymbol{G}_{ij} = 2(\omega_0 - \omega)^{-1} g_i g_j. 
\label{eq:GarbN}
\end{equation}
The matrix $\boldsymbol{G}$ is a dyadic and thus has a unique non-null eigenvalue which is given by $2\Lambda_{\!{_N}}$.  
Considering the orthogonal matrix $\boldsymbol{O}$ that diagonalizes $\boldsymbol{G}$, {i.e.}, 
$\boldsymbol{G}_O=\boldsymbol{O}\boldsymbol{G}\boldsymbol{O}^{\top}=\text{Diag}\bigl[2\Lambda_{\!{_N}},0,\dots,0\bigr]$, 
then one is able to write $\boldsymbol{H}_O=\boldsymbol{G}_O\oplus\boldsymbol{G}_O$ as the matrix with the eigenvalues of $\boldsymbol{H}$.

The metaplectic operator $\hR(t)$ is associated to the symplectic matrix $\boldsymbol{S}=e^{J\boldsymbol{H}t}$ with
\begin{equation}
J=\begin{pmatrix}\boldsymbol{0}_{N} & \boldsymbol{I}_N\\
-\boldsymbol{I}_N & \boldsymbol{0}_{N}
\end{pmatrix}, \label{eq:J}
\end{equation}
where $\boldsymbol{0}_{N}$ denotes the $N \times N$ null matrix and $\boldsymbol{I}_N$ the $N \times N$ identity. 
Under the orthogonal transformation $\boldsymbol{O} \oplus \boldsymbol{O}$, 
the matrix $\boldsymbol{S}_O = \boldsymbol{O} \oplus \boldsymbol{O} \, \boldsymbol{S} (\boldsymbol{O} \oplus \boldsymbol{O})^\top $ 
acquires the form 
\begin{equation}
\boldsymbol{S}_O=\left(\begin{array}{cc}
\cos\left(\boldsymbol{G}_Ot\right) & \sin\left(\boldsymbol{G}_Ot\right)\\
-\sin\left(\boldsymbol{G}_Ot\right) & \cos\left(\boldsymbol{G}_Ot\right)
\end{array}\right).\label{eq:SOarbN}
\end{equation}
In addition, the {\it Cayley parametrization} of $\boldsymbol{S}$ defined by 
\begin{equation}
\boldsymbol{C}= J \left(\boldsymbol{I}_{2N} - \boldsymbol{S}\right)\left(\boldsymbol{I}_{2N} + \boldsymbol{S} \right)^{-1} \label{cayley}
\end{equation}
under the same transformation reads 
\begin{equation}
\boldsymbol{C}_O= J \left(\boldsymbol{I}_{2N}-\boldsymbol{S}_O\right)\left(\boldsymbol{I}_{2N} + \boldsymbol{S}_O\right)^{-1}. 
\end{equation}

Using the above definitions, considering $\hrh_E$ a Gaussian state with covariance matrix $\boldsymbol{\mathcal{V}}$ and null mean values, 
we resort to the Wigner representation to express an average value as \cite{Wigner1932,Nicacio2017,Gosson2018} 
$\Tr\bigl[\hrh\hA\bigr]=\int \! dx \, W\!(x)\, A(x)$, with $W(x)$ being the Wigner function of $\hrh_E$ and $A(x)$ is the {\it center symbol} of $\hat A$. Then one finds
\begin{align}
r_{\!{_N}}(t)&=\Tr\left[\hrh_E\hR(t)\right] \nonumber \\
    {}&=\frac{e^{i\Lambda_{\!{_N}}t}}{\sqrt{{\det\left(\boldsymbol{S}+\boldsymbol{I}_{2N}\right)}\det\left(\frac{1}{2}\boldsymbol{I}_{2N}+i\boldsymbol{\mathcal{V}}\boldsymbol{C}\right)}}.\label{eq:degrN}
\end{align}

If the environmental state $\hrh_E$ is an $N$-mode thermal state, then its covariance matrix becomes 
$\boldsymbol{\mathcal{V}} = \left(\bar{n}+\frac{1}{2}\right)\boldsymbol{I}_{2N}$. 
Inserting this into (\ref{eq:degrN}), and using the fact that $\det(\boldsymbol{O}\oplus\boldsymbol{O}) = 1$, 
since $\boldsymbol{O}$ is orthogonal, Eq.~(\ref{eq:degrN}) becomes 
\[
r_{\!{_N}}(t) = \frac{e^{i\Lambda_{\!{_N}}t}}{\sqrt{{\det\left(\boldsymbol{S_O}+\boldsymbol{I}_{2N}\right)}\det\left[\frac{1}{2}\boldsymbol{I}_{2N}
+ i (\bar{n}+\frac{1}{2})\boldsymbol{C_O}\right]}}, 
\]
which leads directly to Eq.~(\ref{eq:rNdeg}). 

As a final comment, we have implicitly assumed that 
$\det\left(\boldsymbol{S}+\boldsymbol{I}_{2N}\right) \ne 0$ in (\ref{eq:degrN}). 
However, if this happens to not be the case at some instant of time, then such a choice does not invalidate Eq.~(\ref{eq:rNdeg}) but its derivation follows a
complementary procedure based on symplectic Fourier transformations. This is discussed in depth in Refs. \cite{Nicacio2017,Gosson2018}.

\bibliography{refs}

\begin{thebibliography}{50}%
\makeatletter
\providecommand \@ifxundefined [1]{%
 \@ifx{#1\undefined}
}%
\providecommand \@ifnum [1]{%
 \ifnum #1\expandafter \@firstoftwo
 \else \expandafter \@secondoftwo
 \fi
}%
\providecommand \@ifx [1]{%
 \ifx #1\expandafter \@firstoftwo
 \else \expandafter \@secondoftwo
 \fi
}%
\providecommand \natexlab [1]{#1}%
\providecommand \enquote  [1]{``#1''}%
\providecommand \bibnamefont  [1]{#1}%
\providecommand \bibfnamefont [1]{#1}%
\providecommand \citenamefont [1]{#1}%
\providecommand \href@noop [0]{\@secondoftwo}%
\providecommand \href [0]{\begingroup \@sanitize@url \@href}%
\providecommand \@href[1]{\@@startlink{#1}\@@href}%
\providecommand \@@href[1]{\endgroup#1\@@endlink}%
\providecommand \@sanitize@url [0]{\catcode `\\12\catcode `\$12\catcode
  `\&12\catcode `\#12\catcode `\^12\catcode `\_12\catcode `\%12\relax}%
\providecommand \@@startlink[1]{}%
\providecommand \@@endlink[0]{}%
\providecommand \url  [0]{\begingroup\@sanitize@url \@url }%
\providecommand \@url [1]{\endgroup\@href {#1}{\urlprefix }}%
\providecommand \urlprefix  [0]{URL }%
\providecommand \Eprint [0]{\href }%
\providecommand \doibase [0]{http://dx.doi.org/}%
\providecommand \selectlanguage [0]{\@gobble}%
\providecommand \bibinfo  [0]{\@secondoftwo}%
\providecommand \bibfield  [0]{\@secondoftwo}%
\providecommand \translation [1]{[#1]}%
\providecommand \BibitemOpen [0]{}%
\providecommand \bibitemStop [0]{}%
\providecommand \bibitemNoStop [0]{.\EOS\space}%
\providecommand \EOS [0]{\spacefactor3000\relax}%
\providecommand \BibitemShut  [1]{\csname bibitem#1\endcsname}%
\let\auto@bib@innerbib\@empty
\bibitem [{\citenamefont {Schlosshauer}(2008)}]{Schlosshauer2008}%
  \BibitemOpen
  \bibfield  {author} {\bibinfo {author} {\bibfnamefont {M.~A.}\ \bibnamefont
  {Schlosshauer}},\ }\href@noop {} {\emph {\bibinfo {title} {Decoherence and
  the Quantum-to-Classical Transition}}}\ (\bibinfo  {publisher} {Springer},\
  \bibinfo {year} {2008})\BibitemShut {NoStop}%
\bibitem [{\citenamefont {Plenio}\ and\ \citenamefont
  {Huelga}(2008)}]{Plenio2008}%
  \BibitemOpen
  \bibfield  {author} {\bibinfo {author} {\bibfnamefont {M.~B.}\ \bibnamefont
  {Plenio}}\ and\ \bibinfo {author} {\bibfnamefont {S.~F.}\ \bibnamefont
  {Huelga}},\ }\href {\doibase 10.1088/1367-2630/10/11/113019} {\bibfield
  {journal} {\bibinfo  {journal} {New J. Phys.}\ }\textbf {\bibinfo {volume}
  {10}},\ \bibinfo {pages} {113019} (\bibinfo {year} {2008})}\BibitemShut
  {NoStop}%
\bibitem [{\citenamefont {Sinayskiy}\ \emph {et~al.}(2012)\citenamefont
  {Sinayskiy}, \citenamefont {Marais}, \citenamefont {Petruccione},\ and\
  \citenamefont {Ekert}}]{Sinayskiy2012}%
  \BibitemOpen
  \bibfield  {author} {\bibinfo {author} {\bibfnamefont {I.}~\bibnamefont
  {Sinayskiy}}, \bibinfo {author} {\bibfnamefont {A.}~\bibnamefont {Marais}},
  \bibinfo {author} {\bibfnamefont {F.}~\bibnamefont {Petruccione}}, \ and\
  \bibinfo {author} {\bibfnamefont {A.}~\bibnamefont {Ekert}},\ }\href
  {\doibase 10.1103/physrevlett.108.020602} {\bibfield  {journal} {\bibinfo
  {journal} {Phys. Rev. Lett.}\ }\textbf {\bibinfo {volume} {108}},\ \bibinfo
  {pages} {020602} (\bibinfo {year} {2012})}\BibitemShut {NoStop}%
\bibitem [{\citenamefont {Marais}\ \emph {et~al.}(2013)\citenamefont {Marais},
  \citenamefont {Sinayskiy}, \citenamefont {Kay}, \citenamefont {Petruccione},\
  and\ \citenamefont {Ekert}}]{Marais2013}%
  \BibitemOpen
  \bibfield  {author} {\bibinfo {author} {\bibfnamefont {A.}~\bibnamefont
  {Marais}}, \bibinfo {author} {\bibfnamefont {I.}~\bibnamefont {Sinayskiy}},
  \bibinfo {author} {\bibfnamefont {A.}~\bibnamefont {Kay}}, \bibinfo {author}
  {\bibfnamefont {F.}~\bibnamefont {Petruccione}}, \ and\ \bibinfo {author}
  {\bibfnamefont {A.}~\bibnamefont {Ekert}},\ }\href {\doibase
  10.1088/1367-2630/15/1/013038} {\bibfield  {journal} {\bibinfo  {journal}
  {New J. Phys.}\ }\textbf {\bibinfo {volume} {15}},\ \bibinfo {pages} {013038}
  (\bibinfo {year} {2013})}\BibitemShut {NoStop}%
\bibitem [{\citenamefont {Poyatos}\ \emph {et~al.}(1996)\citenamefont
  {Poyatos}, \citenamefont {Cirac},\ and\ \citenamefont
  {Zoller}}]{Poyatos1996}%
  \BibitemOpen
  \bibfield  {author} {\bibinfo {author} {\bibfnamefont {J.~F.}\ \bibnamefont
  {Poyatos}}, \bibinfo {author} {\bibfnamefont {J.~I.}\ \bibnamefont {Cirac}},
  \ and\ \bibinfo {author} {\bibfnamefont {P.}~\bibnamefont {Zoller}},\ }\href
  {\doibase 10.1103/physrevlett.77.4728} {\bibfield  {journal} {\bibinfo
  {journal} {Phys. Rev. Lett.}\ }\textbf {\bibinfo {volume} {77}},\ \bibinfo
  {pages} {4728} (\bibinfo {year} {1996})}\BibitemShut {NoStop}%
\bibitem [{\citenamefont {Carvalho}\ \emph {et~al.}(2001)\citenamefont
  {Carvalho}, \citenamefont {Milman}, \citenamefont {de~Matos~Filho},\ and\
  \citenamefont {Davidovich}}]{Carvalho2001}%
  \BibitemOpen
  \bibfield  {author} {\bibinfo {author} {\bibfnamefont {A.~R.~R.}\
  \bibnamefont {Carvalho}}, \bibinfo {author} {\bibfnamefont {P.}~\bibnamefont
  {Milman}}, \bibinfo {author} {\bibfnamefont {R.~L.}\ \bibnamefont
  {de~Matos~Filho}}, \ and\ \bibinfo {author} {\bibfnamefont {L.}~\bibnamefont
  {Davidovich}},\ }\href {\doibase 10.1103/physrevlett.86.4988} {\bibfield
  {journal} {\bibinfo  {journal} {Phys. Rev. Lett.}\ }\textbf {\bibinfo
  {volume} {86}},\ \bibinfo {pages} {4988} (\bibinfo {year}
  {2001})}\BibitemShut {NoStop}%
\bibitem [{\citenamefont {Roszak}\ \emph {et~al.}(2015)\citenamefont {Roszak},
  \citenamefont {Filip},\ and\ \citenamefont {Novotn{\'{y}}}}]{Roszak2015}%
  \BibitemOpen
  \bibfield  {author} {\bibinfo {author} {\bibfnamefont {K.}~\bibnamefont
  {Roszak}}, \bibinfo {author} {\bibfnamefont {R.}~\bibnamefont {Filip}}, \
  and\ \bibinfo {author} {\bibfnamefont {T.}~\bibnamefont {Novotn{\'{y}}}},\
  }\href {\doibase 10.1038/srep09796} {\bibfield  {journal} {\bibinfo
  {journal} {Sci. Rep.}\ }\textbf {\bibinfo {volume} {5}},\ \bibinfo {pages}
  {9796} (\bibinfo {year} {2015})}\BibitemShut {NoStop}%
\bibitem [{\citenamefont {Brune}\ \emph {et~al.}(1996)\citenamefont {Brune},
  \citenamefont {Hagley}, \citenamefont {Dreyer}, \citenamefont
  {Ma{\^{\i}}tre}, \citenamefont {Maali}, \citenamefont {Wunderlich},
  \citenamefont {Raimond},\ and\ \citenamefont {Haroche}}]{Brune1996}%
  \BibitemOpen
  \bibfield  {author} {\bibinfo {author} {\bibfnamefont {M.}~\bibnamefont
  {Brune}}, \bibinfo {author} {\bibfnamefont {E.}~\bibnamefont {Hagley}},
  \bibinfo {author} {\bibfnamefont {J.}~\bibnamefont {Dreyer}}, \bibinfo
  {author} {\bibfnamefont {X.}~\bibnamefont {Ma{\^{\i}}tre}}, \bibinfo {author}
  {\bibfnamefont {A.}~\bibnamefont {Maali}}, \bibinfo {author} {\bibfnamefont
  {C.}~\bibnamefont {Wunderlich}}, \bibinfo {author} {\bibfnamefont {J.~M.}\
  \bibnamefont {Raimond}}, \ and\ \bibinfo {author} {\bibfnamefont
  {S.}~\bibnamefont {Haroche}},\ }\href {\doibase 10.1103/physrevlett.77.4887}
  {\bibfield  {journal} {\bibinfo  {journal} {Phys. Rev. Lett.}\ }\textbf
  {\bibinfo {volume} {77}},\ \bibinfo {pages} {4887} (\bibinfo {year}
  {1996})}\BibitemShut {NoStop}%
\bibitem [{\citenamefont {Myatt}\ \emph {et~al.}(2000)\citenamefont {Myatt},
  \citenamefont {King}, \citenamefont {Turchette}, \citenamefont {Sackett},
  \citenamefont {Kielpinski}, \citenamefont {Itano}, \citenamefont {Monroe},\
  and\ \citenamefont {Wineland}}]{Myatt2000}%
  \BibitemOpen
  \bibfield  {author} {\bibinfo {author} {\bibfnamefont {C.~J.}\ \bibnamefont
  {Myatt}}, \bibinfo {author} {\bibfnamefont {B.~E.}\ \bibnamefont {King}},
  \bibinfo {author} {\bibfnamefont {Q.~A.}\ \bibnamefont {Turchette}}, \bibinfo
  {author} {\bibfnamefont {C.~A.}\ \bibnamefont {Sackett}}, \bibinfo {author}
  {\bibfnamefont {D.}~\bibnamefont {Kielpinski}}, \bibinfo {author}
  {\bibfnamefont {W.~M.}\ \bibnamefont {Itano}}, \bibinfo {author}
  {\bibfnamefont {C.}~\bibnamefont {Monroe}}, \ and\ \bibinfo {author}
  {\bibfnamefont {D.~J.}\ \bibnamefont {Wineland}},\ }\href {\doibase
  10.1038/35002001} {\bibfield  {journal} {\bibinfo  {journal} {Nature}\
  }\textbf {\bibinfo {volume} {403}},\ \bibinfo {pages} {269} (\bibinfo {year}
  {2000})}\BibitemShut {NoStop}%
\bibitem [{\citenamefont {Bertet}\ \emph {et~al.}(2005)\citenamefont {Bertet},
  \citenamefont {Chiorescu}, \citenamefont {Burkard}, \citenamefont {Semba},
  \citenamefont {Harmans}, \citenamefont {DiVincenzo},\ and\ \citenamefont
  {Mooij}}]{Bertet2005}%
  \BibitemOpen
  \bibfield  {author} {\bibinfo {author} {\bibfnamefont {P.}~\bibnamefont
  {Bertet}}, \bibinfo {author} {\bibfnamefont {I.}~\bibnamefont {Chiorescu}},
  \bibinfo {author} {\bibfnamefont {G.}~\bibnamefont {Burkard}}, \bibinfo
  {author} {\bibfnamefont {K.}~\bibnamefont {Semba}}, \bibinfo {author}
  {\bibfnamefont {C.~J. P.~M.}\ \bibnamefont {Harmans}}, \bibinfo {author}
  {\bibfnamefont {D.~P.}\ \bibnamefont {DiVincenzo}}, \ and\ \bibinfo {author}
  {\bibfnamefont {J.~E.}\ \bibnamefont {Mooij}},\ }\href {\doibase
  10.1103/physrevlett.95.257002} {\bibfield  {journal} {\bibinfo  {journal}
  {Phys. Rev. Lett.}\ }\textbf {\bibinfo {volume} {95}},\ \bibinfo {pages}
  {257002} (\bibinfo {year} {2005})}\BibitemShut {NoStop}%
\bibitem [{\citenamefont {Buono}\ \emph {et~al.}(2012)\citenamefont {Buono},
  \citenamefont {Nocerino}, \citenamefont {Porzio},\ and\ \citenamefont
  {Solimeno}}]{Buono2012}%
  \BibitemOpen
  \bibfield  {author} {\bibinfo {author} {\bibfnamefont {D.}~\bibnamefont
  {Buono}}, \bibinfo {author} {\bibfnamefont {G.}~\bibnamefont {Nocerino}},
  \bibinfo {author} {\bibfnamefont {A.}~\bibnamefont {Porzio}}, \ and\ \bibinfo
  {author} {\bibfnamefont {S.}~\bibnamefont {Solimeno}},\ }\href {\doibase
  10.1103/physreva.86.042308} {\bibfield  {journal} {\bibinfo  {journal} {Phys.
  Rev. A}\ }\textbf {\bibinfo {volume} {86}},\ \bibinfo {pages} {042308}
  (\bibinfo {year} {2012})}\BibitemShut {NoStop}%
\bibitem [{\citenamefont {Schneider}\ \emph {et~al.}(2014)\citenamefont
  {Schneider}, \citenamefont {Singh}, \citenamefont {Venstra}, \citenamefont
  {Meerwaldt},\ and\ \citenamefont {Steele}}]{Schneider2014}%
  \BibitemOpen
  \bibfield  {author} {\bibinfo {author} {\bibfnamefont {B.~H.}\ \bibnamefont
  {Schneider}}, \bibinfo {author} {\bibfnamefont {V.}~\bibnamefont {Singh}},
  \bibinfo {author} {\bibfnamefont {W.~J.}\ \bibnamefont {Venstra}}, \bibinfo
  {author} {\bibfnamefont {H.~B.}\ \bibnamefont {Meerwaldt}}, \ and\ \bibinfo
  {author} {\bibfnamefont {G.~A.}\ \bibnamefont {Steele}},\ }\href {\doibase
  10.1038/ncomms6819} {\bibfield  {journal} {\bibinfo  {journal} {Nat.
  Commun.}\ }\textbf {\bibinfo {volume} {5}},\ \bibinfo {pages} {5819}
  (\bibinfo {year} {2014})}\BibitemShut {NoStop}%
\bibitem [{\citenamefont {Zurek}(1991)}]{Zurek1991}%
  \BibitemOpen
  \bibfield  {author} {\bibinfo {author} {\bibfnamefont {W.~H.}\ \bibnamefont
  {Zurek}},\ }\href {\doibase 10.1063/1.881293} {\bibfield  {journal} {\bibinfo
   {journal} {Phys. Today}\ }\textbf {\bibinfo {volume} {44}},\ \bibinfo
  {pages} {36} (\bibinfo {year} {1991})}\BibitemShut {NoStop}%
\bibitem [{\citenamefont {Leggett}\ \emph {et~al.}(1987)\citenamefont
  {Leggett}, \citenamefont {Chakravarty}, \citenamefont {Dorsey}, \citenamefont
  {Fisher}, \citenamefont {Garg},\ and\ \citenamefont {Zwerger}}]{Leggett1987}%
  \BibitemOpen
  \bibfield  {author} {\bibinfo {author} {\bibfnamefont {A.~J.}\ \bibnamefont
  {Leggett}}, \bibinfo {author} {\bibfnamefont {S.}~\bibnamefont
  {Chakravarty}}, \bibinfo {author} {\bibfnamefont {A.~T.}\ \bibnamefont
  {Dorsey}}, \bibinfo {author} {\bibfnamefont {M.~P.~A.}\ \bibnamefont
  {Fisher}}, \bibinfo {author} {\bibfnamefont {A.}~\bibnamefont {Garg}}, \ and\
  \bibinfo {author} {\bibfnamefont {W.}~\bibnamefont {Zwerger}},\ }\href
  {\doibase 10.1103/revmodphys.59.1} {\bibfield  {journal} {\bibinfo  {journal}
  {Rev. Mod. Phys.}\ }\textbf {\bibinfo {volume} {59}},\ \bibinfo {pages} {1}
  (\bibinfo {year} {1987})}\BibitemShut {NoStop}%
\bibitem [{\citenamefont {Prokof'ev}\ and\ \citenamefont
  {Stamp}(2000)}]{Proko2000}%
  \BibitemOpen
  \bibfield  {author} {\bibinfo {author} {\bibfnamefont {N.~V.}\ \bibnamefont
  {Prokof'ev}}\ and\ \bibinfo {author} {\bibfnamefont {P.~C.~E.}\ \bibnamefont
  {Stamp}},\ }\href {\doibase 10.1088/0034-4885/63/4/204} {\bibfield  {journal}
  {\bibinfo  {journal} {Rep. Prog. Phys.}\ }\textbf {\bibinfo {volume} {63}},\
  \bibinfo {pages} {669} (\bibinfo {year} {2000})}\BibitemShut {NoStop}%
\bibitem [{\citenamefont {Turchette}\ \emph {et~al.}(2000)\citenamefont
  {Turchette}, \citenamefont {Myatt}, \citenamefont {King}, \citenamefont
  {Sackett}, \citenamefont {Kielpinski}, \citenamefont {Itano}, \citenamefont
  {Monroe},\ and\ \citenamefont {Wineland}}]{Turchette2000}%
  \BibitemOpen
  \bibfield  {author} {\bibinfo {author} {\bibfnamefont {Q.~A.}\ \bibnamefont
  {Turchette}}, \bibinfo {author} {\bibfnamefont {C.~J.}\ \bibnamefont
  {Myatt}}, \bibinfo {author} {\bibfnamefont {B.~E.}\ \bibnamefont {King}},
  \bibinfo {author} {\bibfnamefont {C.~A.}\ \bibnamefont {Sackett}}, \bibinfo
  {author} {\bibfnamefont {D.}~\bibnamefont {Kielpinski}}, \bibinfo {author}
  {\bibfnamefont {W.~M.}\ \bibnamefont {Itano}}, \bibinfo {author}
  {\bibfnamefont {C.}~\bibnamefont {Monroe}}, \ and\ \bibinfo {author}
  {\bibfnamefont {D.~J.}\ \bibnamefont {Wineland}},\ }\href {\doibase
  10.1103/physreva.62.053807} {\bibfield  {journal} {\bibinfo  {journal} {Phys.
  Rev. A}\ }\textbf {\bibinfo {volume} {62}},\ \bibinfo {pages} {053807}
  (\bibinfo {year} {2000})}\BibitemShut {NoStop}%
\bibitem [{\citenamefont {Liu}\ \emph {et~al.}(2018)\citenamefont {Liu},
  \citenamefont {Lyyra}, \citenamefont {Sun}, \citenamefont {Liu},
  \citenamefont {Li}, \citenamefont {Guo}, \citenamefont {Maniscalco},\ and\
  \citenamefont {Piilo}}]{Liu2018}%
  \BibitemOpen
  \bibfield  {author} {\bibinfo {author} {\bibfnamefont {Z.-D.}\ \bibnamefont
  {Liu}}, \bibinfo {author} {\bibfnamefont {H.}~\bibnamefont {Lyyra}}, \bibinfo
  {author} {\bibfnamefont {Y.-N.}\ \bibnamefont {Sun}}, \bibinfo {author}
  {\bibfnamefont {B.-H.}\ \bibnamefont {Liu}}, \bibinfo {author} {\bibfnamefont
  {C.-F.}\ \bibnamefont {Li}}, \bibinfo {author} {\bibfnamefont {G.-C.}\
  \bibnamefont {Guo}}, \bibinfo {author} {\bibfnamefont {S.}~\bibnamefont
  {Maniscalco}}, \ and\ \bibinfo {author} {\bibfnamefont {J.}~\bibnamefont
  {Piilo}},\ }\href {\doibase 10.1038/s41467-018-05817-x} {\bibfield  {journal}
  {\bibinfo  {journal} {Nat. Commun.}\ }\textbf {\bibinfo {volume} {9}},\
  \bibinfo {pages} {3453} (\bibinfo {year} {2018})}\BibitemShut {NoStop}%
\bibitem [{\citenamefont {Jaynes}\ and\ \citenamefont
  {Cummings}(1963)}]{Jaynes1963}%
  \BibitemOpen
  \bibfield  {author} {\bibinfo {author} {\bibfnamefont {E.~T.}\ \bibnamefont
  {Jaynes}}\ and\ \bibinfo {author} {\bibfnamefont {F.~W.}\ \bibnamefont
  {Cummings}},\ }\href {\doibase 10.1109/PROC.1963.1664} {\bibfield  {journal}
  {\bibinfo  {journal} {Proc. IEEE}\ }\textbf {\bibinfo {volume} {51}},\
  \bibinfo {pages} {89} (\bibinfo {year} {1963})}\BibitemShut {NoStop}%
\bibitem [{\citenamefont {Brune}\ \emph {et~al.}(1992)\citenamefont {Brune},
  \citenamefont {Haroche}, \citenamefont {Raimond}, \citenamefont
  {Davidovich},\ and\ \citenamefont {Zagury}}]{Brune1992}%
  \BibitemOpen
  \bibfield  {author} {\bibinfo {author} {\bibfnamefont {M.}~\bibnamefont
  {Brune}}, \bibinfo {author} {\bibfnamefont {S.}~\bibnamefont {Haroche}},
  \bibinfo {author} {\bibfnamefont {J.~M.}\ \bibnamefont {Raimond}}, \bibinfo
  {author} {\bibfnamefont {L.}~\bibnamefont {Davidovich}}, \ and\ \bibinfo
  {author} {\bibfnamefont {N.}~\bibnamefont {Zagury}},\ }\href {\doibase
  10.1103/physreva.45.5193} {\bibfield  {journal} {\bibinfo  {journal} {Phys.
  Rev. A}\ }\textbf {\bibinfo {volume} {45}},\ \bibinfo {pages} {5193}
  (\bibinfo {year} {1992})}\BibitemShut {NoStop}%
\bibitem [{\citenamefont {Blais}\ \emph {et~al.}(2004)\citenamefont {Blais},
  \citenamefont {Huang}, \citenamefont {Wallraff}, \citenamefont {Girvin},\
  and\ \citenamefont {Schoelkopf}}]{Blais2004}%
  \BibitemOpen
  \bibfield  {author} {\bibinfo {author} {\bibfnamefont {A.}~\bibnamefont
  {Blais}}, \bibinfo {author} {\bibfnamefont {R.-S.}\ \bibnamefont {Huang}},
  \bibinfo {author} {\bibfnamefont {A.}~\bibnamefont {Wallraff}}, \bibinfo
  {author} {\bibfnamefont {S.~M.}\ \bibnamefont {Girvin}}, \ and\ \bibinfo
  {author} {\bibfnamefont {R.~J.}\ \bibnamefont {Schoelkopf}},\ }\href
  {\doibase 10.1103/physreva.69.062320} {\bibfield  {journal} {\bibinfo
  {journal} {Phys. Rev. A}\ }\textbf {\bibinfo {volume} {69}},\ \bibinfo
  {pages} {062320} (\bibinfo {year} {2004})}\BibitemShut {NoStop}%
\bibitem [{\citenamefont {Carmichael}(2013)}]{Carmichael2013}%
  \BibitemOpen
  \bibfield  {author} {\bibinfo {author} {\bibfnamefont {H.}~\bibnamefont
  {Carmichael}},\ }\href@noop {} {\emph {\bibinfo {title} {An Open Systems
  Approach to Quantum Optics}}}\ (\bibinfo  {publisher} {Springer},\ \bibinfo
  {year} {2013})\BibitemShut {NoStop}%
\bibitem [{\citenamefont {Leandro}\ and\ \citenamefont
  {Semi{\~{a}}o}(2009)}]{Leandro2009}%
  \BibitemOpen
  \bibfield  {author} {\bibinfo {author} {\bibfnamefont {J.~F.}\ \bibnamefont
  {Leandro}}\ and\ \bibinfo {author} {\bibfnamefont {F.~L.}\ \bibnamefont
  {Semi{\~{a}}o}},\ }\href {\doibase 10.1016/j.optcom.2009.08.059} {\bibfield
  {journal} {\bibinfo  {journal} {Opt. Commun.}\ }\textbf {\bibinfo {volume}
  {282}},\ \bibinfo {pages} {4736} (\bibinfo {year} {2009})}\BibitemShut
  {NoStop}%
\bibitem [{\citenamefont {Breuer}\ \emph {et~al.}(1999)\citenamefont {Breuer},
  \citenamefont {Kappler},\ and\ \citenamefont {Petruccione}}]{Breuer1999}%
  \BibitemOpen
  \bibfield  {author} {\bibinfo {author} {\bibfnamefont {H.-P.}\ \bibnamefont
  {Breuer}}, \bibinfo {author} {\bibfnamefont {B.}~\bibnamefont {Kappler}}, \
  and\ \bibinfo {author} {\bibfnamefont {F.}~\bibnamefont {Petruccione}},\
  }\href {\doibase 10.1103/physreva.59.1633} {\bibfield  {journal} {\bibinfo
  {journal} {Phys. Rev. A}\ }\textbf {\bibinfo {volume} {59}},\ \bibinfo
  {pages} {1633} (\bibinfo {year} {1999})}\BibitemShut {NoStop}%
\bibitem [{\citenamefont {Vacchini}\ and\ \citenamefont
  {Breuer}(2010)}]{Vacchini2010}%
  \BibitemOpen
  \bibfield  {author} {\bibinfo {author} {\bibfnamefont {B.}~\bibnamefont
  {Vacchini}}\ and\ \bibinfo {author} {\bibfnamefont {H.-P.}\ \bibnamefont
  {Breuer}},\ }\href {\doibase 10.1103/physreva.81.042103} {\bibfield
  {journal} {\bibinfo  {journal} {Phys. Rev. A}\ }\textbf {\bibinfo {volume}
  {81}},\ \bibinfo {pages} {042103} (\bibinfo {year} {2010})}\BibitemShut
  {NoStop}%
\bibitem [{\citenamefont {Garraway}(1997)}]{Garraway1997}%
  \BibitemOpen
  \bibfield  {author} {\bibinfo {author} {\bibfnamefont {B.~M.}\ \bibnamefont
  {Garraway}},\ }\href {\doibase 10.1103/physreva.55.2290} {\bibfield
  {journal} {\bibinfo  {journal} {Phys. Rev. A}\ }\textbf {\bibinfo {volume}
  {55}},\ \bibinfo {pages} {2290} (\bibinfo {year} {1997})}\BibitemShut
  {NoStop}%
\bibitem [{\citenamefont {Mazzola}\ \emph {et~al.}(2009)\citenamefont
  {Mazzola}, \citenamefont {Maniscalco}, \citenamefont {Piilo}, \citenamefont
  {Suominen},\ and\ \citenamefont {Garraway}}]{Mazzola2009}%
  \BibitemOpen
  \bibfield  {author} {\bibinfo {author} {\bibfnamefont {L.}~\bibnamefont
  {Mazzola}}, \bibinfo {author} {\bibfnamefont {S.}~\bibnamefont {Maniscalco}},
  \bibinfo {author} {\bibfnamefont {J.}~\bibnamefont {Piilo}}, \bibinfo
  {author} {\bibfnamefont {K.-A.}\ \bibnamefont {Suominen}}, \ and\ \bibinfo
  {author} {\bibfnamefont {B.~M.}\ \bibnamefont {Garraway}},\ }\href {\doibase
  10.1103/physreva.80.012104} {\bibfield  {journal} {\bibinfo  {journal} {Phys.
  Rev. A}\ }\textbf {\bibinfo {volume} {80}},\ \bibinfo {pages} {012104}
  (\bibinfo {year} {2009})}\BibitemShut {NoStop}%
\bibitem [{\citenamefont {Magnus}(1954)}]{Magnus1954}%
  \BibitemOpen
  \bibfield  {author} {\bibinfo {author} {\bibfnamefont {W.}~\bibnamefont
  {Magnus}},\ }\href {\doibase 10.1002/cpa.3160070404} {\bibfield  {journal}
  {\bibinfo  {journal} {Commun. Pure Appl. Math.}\ }\textbf {\bibinfo {volume}
  {7}},\ \bibinfo {pages} {649} (\bibinfo {year} {1954})}\BibitemShut {NoStop}%
\bibitem [{\citenamefont {Baumgratz}\ \emph {et~al.}(2014)\citenamefont
  {Baumgratz}, \citenamefont {Cramer},\ and\ \citenamefont
  {Plenio}}]{Baumgratz2013}%
  \BibitemOpen
  \bibfield  {author} {\bibinfo {author} {\bibfnamefont {T.}~\bibnamefont
  {Baumgratz}}, \bibinfo {author} {\bibfnamefont {M.}~\bibnamefont {Cramer}}, \
  and\ \bibinfo {author} {\bibfnamefont {M.~B.}\ \bibnamefont {Plenio}},\
  }\href {\doibase 10.1103/PhysRevLett.113.140401} {\bibfield  {journal}
  {\bibinfo  {journal} {Phys. Rev. Lett.}\ }\textbf {\bibinfo {volume} {113}},\
  \bibinfo {pages} {140401} (\bibinfo {year} {2014})}\BibitemShut {NoStop}%
\bibitem [{\citenamefont {Glauber}(1963)}]{Glauber1963}%
  \BibitemOpen
  \bibfield  {author} {\bibinfo {author} {\bibfnamefont {R.~J.}\ \bibnamefont
  {Glauber}},\ }\href {\doibase 10.1103/physrev.131.2766} {\bibfield  {journal}
  {\bibinfo  {journal} {Phys. Rev.}\ }\textbf {\bibinfo {volume} {131}},\
  \bibinfo {pages} {2766} (\bibinfo {year} {1963})}\BibitemShut {NoStop}%
\bibitem [{\citenamefont {Sudarshan}(1963)}]{Sudarshan1963}%
  \BibitemOpen
  \bibfield  {author} {\bibinfo {author} {\bibfnamefont {E.~C.~G.}\
  \bibnamefont {Sudarshan}},\ }\href {\doibase 10.1103/physrevlett.10.277}
  {\bibfield  {journal} {\bibinfo  {journal} {Phys. Rev. Lett.}\ }\textbf
  {\bibinfo {volume} {10}},\ \bibinfo {pages} {277} (\bibinfo {year}
  {1963})}\BibitemShut {NoStop}%
\bibitem [{\citenamefont {Breuer}\ \emph {et~al.}(2009)\citenamefont {Breuer},
  \citenamefont {Laine},\ and\ \citenamefont {Piilo}}]{Breuer2009}%
  \BibitemOpen
  \bibfield  {author} {\bibinfo {author} {\bibfnamefont {H.-P.}\ \bibnamefont
  {Breuer}}, \bibinfo {author} {\bibfnamefont {E.-M.}\ \bibnamefont {Laine}}, \
  and\ \bibinfo {author} {\bibfnamefont {J.}~\bibnamefont {Piilo}},\ }\href
  {\doibase 10.1103/physrevlett.103.210401} {\bibfield  {journal} {\bibinfo
  {journal} {Phys. Rev. Lett.}\ }\textbf {\bibinfo {volume} {103}},\ \bibinfo
  {pages} {210401} (\bibinfo {year} {2009})}\BibitemShut {NoStop}%
\bibitem [{\citenamefont {Wi{\ss}mann}\ \emph {et~al.}(2012)\citenamefont
  {Wi{\ss}mann}, \citenamefont {Karlsson}, \citenamefont {Laine}, \citenamefont
  {Piilo},\ and\ \citenamefont {Breuer}}]{Wismann2012}%
  \BibitemOpen
  \bibfield  {author} {\bibinfo {author} {\bibfnamefont {S.}~\bibnamefont
  {Wi{\ss}mann}}, \bibinfo {author} {\bibfnamefont {A.}~\bibnamefont
  {Karlsson}}, \bibinfo {author} {\bibfnamefont {E.-M.}\ \bibnamefont {Laine}},
  \bibinfo {author} {\bibfnamefont {J.}~\bibnamefont {Piilo}}, \ and\ \bibinfo
  {author} {\bibfnamefont {H.-P.}\ \bibnamefont {Breuer}},\ }\href {\doibase
  10.1103/physreva.86.062108} {\bibfield  {journal} {\bibinfo  {journal} {Phys.
  Rev. A}\ }\textbf {\bibinfo {volume} {86}},\ \bibinfo {pages} {062108}
  (\bibinfo {year} {2012})}\BibitemShut {NoStop}%
\bibitem [{\citenamefont {Breuer}\ \emph {et~al.}(2016)\citenamefont {Breuer},
  \citenamefont {Laine}, \citenamefont {Piilo},\ and\ \citenamefont
  {Vacchini}}]{Breuer2016}%
  \BibitemOpen
  \bibfield  {author} {\bibinfo {author} {\bibfnamefont {H.-P.}\ \bibnamefont
  {Breuer}}, \bibinfo {author} {\bibfnamefont {E.-M.}\ \bibnamefont {Laine}},
  \bibinfo {author} {\bibfnamefont {J.}~\bibnamefont {Piilo}}, \ and\ \bibinfo
  {author} {\bibfnamefont {B.}~\bibnamefont {Vacchini}},\ }\href {\doibase
  10.1103/revmodphys.88.021002} {\bibfield  {journal} {\bibinfo  {journal}
  {Rev. Mod. Phys.}\ }\textbf {\bibinfo {volume} {88}},\ \bibinfo {pages}
  {021002} (\bibinfo {year} {2016})}\BibitemShut {NoStop}%
\bibitem [{\citenamefont {Uhlig}(2015)}]{Uhlig2015}%
  \BibitemOpen
  \bibfield  {author} {\bibinfo {author} {\bibfnamefont {K.}~\bibnamefont
  {Uhlig}},\ }\href {\doibase 10.1016/j.cryogenics.2014.10.004} {\bibfield
  {journal} {\bibinfo  {journal} {Cryogenics}\ }\textbf {\bibinfo {volume}
  {66}},\ \bibinfo {pages} {6} (\bibinfo {year} {2015})}\BibitemShut {NoStop}%
\bibitem [{\citenamefont {Blais}\ \emph {et~al.}(2007)\citenamefont {Blais},
  \citenamefont {Gambetta}, \citenamefont {Wallraff}, \citenamefont {Schuster},
  \citenamefont {Girvin}, \citenamefont {Devoret},\ and\ \citenamefont
  {Schoelkopf}}]{Blais2007}%
  \BibitemOpen
  \bibfield  {author} {\bibinfo {author} {\bibfnamefont {A.}~\bibnamefont
  {Blais}}, \bibinfo {author} {\bibfnamefont {J.}~\bibnamefont {Gambetta}},
  \bibinfo {author} {\bibfnamefont {A.}~\bibnamefont {Wallraff}}, \bibinfo
  {author} {\bibfnamefont {D.~I.}\ \bibnamefont {Schuster}}, \bibinfo {author}
  {\bibfnamefont {S.~M.}\ \bibnamefont {Girvin}}, \bibinfo {author}
  {\bibfnamefont {M.~H.}\ \bibnamefont {Devoret}}, \ and\ \bibinfo {author}
  {\bibfnamefont {R.~J.}\ \bibnamefont {Schoelkopf}},\ }\href {\doibase
  10.1103/physreva.75.032329} {\bibfield  {journal} {\bibinfo  {journal} {Phys.
  Rev. A}\ }\textbf {\bibinfo {volume} {75}},\ \bibinfo {pages} {032329}
  (\bibinfo {year} {2007})}\BibitemShut {NoStop}%
\bibitem [{\citenamefont {Paik}\ \emph {et~al.}(2011)\citenamefont {Paik},
  \citenamefont {Schuster}, \citenamefont {Bishop}, \citenamefont {Kirchmair},
  \citenamefont {Catelani}, \citenamefont {Sears}, \citenamefont {Johnson},
  \citenamefont {Reagor}, \citenamefont {Frunzio}, \citenamefont {Glazman},
  \citenamefont {Girvin}, \citenamefont {Devoret},\ and\ \citenamefont
  {Schoelkopf}}]{Paik2011}%
  \BibitemOpen
  \bibfield  {author} {\bibinfo {author} {\bibfnamefont {H.}~\bibnamefont
  {Paik}}, \bibinfo {author} {\bibfnamefont {D.~I.}\ \bibnamefont {Schuster}},
  \bibinfo {author} {\bibfnamefont {L.~S.}\ \bibnamefont {Bishop}}, \bibinfo
  {author} {\bibfnamefont {G.}~\bibnamefont {Kirchmair}}, \bibinfo {author}
  {\bibfnamefont {G.}~\bibnamefont {Catelani}}, \bibinfo {author}
  {\bibfnamefont {A.~P.}\ \bibnamefont {Sears}}, \bibinfo {author}
  {\bibfnamefont {B.~R.}\ \bibnamefont {Johnson}}, \bibinfo {author}
  {\bibfnamefont {M.~J.}\ \bibnamefont {Reagor}}, \bibinfo {author}
  {\bibfnamefont {L.}~\bibnamefont {Frunzio}}, \bibinfo {author} {\bibfnamefont
  {L.~I.}\ \bibnamefont {Glazman}}, \bibinfo {author} {\bibfnamefont {S.~M.}\
  \bibnamefont {Girvin}}, \bibinfo {author} {\bibfnamefont {M.~H.}\
  \bibnamefont {Devoret}}, \ and\ \bibinfo {author} {\bibfnamefont {R.~J.}\
  \bibnamefont {Schoelkopf}},\ }\href {\doibase 10.1103/physrevlett.107.240501}
  {\bibfield  {journal} {\bibinfo  {journal} {Phys. Rev. Lett.}\ }\textbf
  {\bibinfo {volume} {107}},\ \bibinfo {pages} {240501} (\bibinfo {year}
  {2011})}\BibitemShut {NoStop}%
\bibitem [{\citenamefont {Gely}\ \emph {et~al.}(2017)\citenamefont {Gely},
  \citenamefont {Parra-Rodriguez}, \citenamefont {Bothner}, \citenamefont
  {Blanter}, \citenamefont {Bosman}, \citenamefont {Solano},\ and\
  \citenamefont {Steele}}]{Gely2017}%
  \BibitemOpen
  \bibfield  {author} {\bibinfo {author} {\bibfnamefont {M.~F.}\ \bibnamefont
  {Gely}}, \bibinfo {author} {\bibfnamefont {A.}~\bibnamefont
  {Parra-Rodriguez}}, \bibinfo {author} {\bibfnamefont {D.}~\bibnamefont
  {Bothner}}, \bibinfo {author} {\bibfnamefont {Y.~M.}\ \bibnamefont
  {Blanter}}, \bibinfo {author} {\bibfnamefont {S.~J.}\ \bibnamefont {Bosman}},
  \bibinfo {author} {\bibfnamefont {E.}~\bibnamefont {Solano}}, \ and\ \bibinfo
  {author} {\bibfnamefont {G.~A.}\ \bibnamefont {Steele}},\ }\href {\doibase
  10.1103/physrevb.95.245115} {\bibfield  {journal} {\bibinfo  {journal} {Phys.
  Rev. B}\ }\textbf {\bibinfo {volume} {95}},\ \bibinfo {pages} {245115}
  (\bibinfo {year} {2017})}\BibitemShut {NoStop}%
\bibitem [{\citenamefont {Malekakhlagh}\ \emph {et~al.}(2017)\citenamefont
  {Malekakhlagh}, \citenamefont {Petrescu},\ and\ \citenamefont
  {T\"ureci}}]{Malekakhlagh2017}%
  \BibitemOpen
  \bibfield  {author} {\bibinfo {author} {\bibfnamefont {M.}~\bibnamefont
  {Malekakhlagh}}, \bibinfo {author} {\bibfnamefont {A.}~\bibnamefont
  {Petrescu}}, \ and\ \bibinfo {author} {\bibfnamefont {H.~E.}\ \bibnamefont
  {T\"ureci}},\ }\href {\doibase 10.1103/physrevlett.119.073601} {\bibfield
  {journal} {\bibinfo  {journal} {Phys. Rev. Lett.}\ }\textbf {\bibinfo
  {volume} {119}},\ \bibinfo {pages} {073601} (\bibinfo {year}
  {2017})}\BibitemShut {NoStop}%
\bibitem [{\citenamefont {Guarnieri}\ \emph {et~al.}(2018)\citenamefont
  {Guarnieri}, \citenamefont {Kol{\'{a}}{\v{r}}},\ and\ \citenamefont
  {Filip}}]{Guarnieri2018}%
  \BibitemOpen
  \bibfield  {author} {\bibinfo {author} {\bibfnamefont {G.}~\bibnamefont
  {Guarnieri}}, \bibinfo {author} {\bibfnamefont {M.}~\bibnamefont
  {Kol{\'{a}}{\v{r}}}}, \ and\ \bibinfo {author} {\bibfnamefont
  {R.}~\bibnamefont {Filip}},\ }\href {\doibase 10.1103/physrevlett.121.070401}
  {\bibfield  {journal} {\bibinfo  {journal} {Phys. Rev. Lett.}\ }\textbf
  {\bibinfo {volume} {121}},\ \bibinfo {pages} {070401} (\bibinfo {year}
  {2018})}\BibitemShut {NoStop}%
\bibitem [{\citenamefont {Wang}\ \emph {et~al.}(2018)\citenamefont {Wang},
  \citenamefont {Wu}, \citenamefont {Cui},\ and\ \citenamefont
  {Wang}}]{Wang2018}%
  \BibitemOpen
  \bibfield  {author} {\bibinfo {author} {\bibfnamefont {Z.}~\bibnamefont
  {Wang}}, \bibinfo {author} {\bibfnamefont {W.}~\bibnamefont {Wu}}, \bibinfo
  {author} {\bibfnamefont {G.}~\bibnamefont {Cui}}, \ and\ \bibinfo {author}
  {\bibfnamefont {J.}~\bibnamefont {Wang}},\ }\href {\doibase
  10.1088/1367-2630/aab03a} {\bibfield  {journal} {\bibinfo  {journal} {New J.
  Phys.}\ }\textbf {\bibinfo {volume} {20}},\ \bibinfo {pages} {033034}
  (\bibinfo {year} {2018})}\BibitemShut {NoStop}%
\bibitem [{\citenamefont {Wang}\ \emph {et~al.}()\citenamefont {Wang},
  \citenamefont {Wu},\ and\ \citenamefont {Wang}}]{Wang2018a}%
  \BibitemOpen
  \bibfield  {author} {\bibinfo {author} {\bibfnamefont {Z.}~\bibnamefont
  {Wang}}, \bibinfo {author} {\bibfnamefont {W.}~\bibnamefont {Wu}}, \ and\
  \bibinfo {author} {\bibfnamefont {J.}~\bibnamefont {Wang}},\ }\href@noop {}
  {\ }\Eprint {http://arxiv.org/abs/arXiv:1812.04799} {arXiv:1812.04799}
  \BibitemShut {NoStop}%
\bibitem [{\citenamefont {Nicacio}\ \emph {et~al.}(2016)\citenamefont
  {Nicacio}, \citenamefont {Paternostro},\ and\ \citenamefont
  {Ferraro}}]{Nicacio2016}%
  \BibitemOpen
  \bibfield  {author} {\bibinfo {author} {\bibfnamefont {F.}~\bibnamefont
  {Nicacio}}, \bibinfo {author} {\bibfnamefont {M.}~\bibnamefont
  {Paternostro}}, \ and\ \bibinfo {author} {\bibfnamefont {A.}~\bibnamefont
  {Ferraro}},\ }\href {\doibase 10.1103/physreva.94.052129} {\bibfield
  {journal} {\bibinfo  {journal} {Phys. Rev. A}\ }\textbf {\bibinfo {volume}
  {94}},\ \bibinfo {pages} {052129} (\bibinfo {year} {2016})}\BibitemShut
  {NoStop}%
\bibitem [{\citenamefont {Barreiro}\ \emph {et~al.}(2011)\citenamefont
  {Barreiro}, \citenamefont {M\"{u}ller}, \citenamefont {Schindler},
  \citenamefont {Nigg}, \citenamefont {Monz}, \citenamefont {Chwalla},
  \citenamefont {Hennrich}, \citenamefont {Roos}, \citenamefont {Zoller},\ and\
  \citenamefont {Blatt}}]{Barreiro2011}%
  \BibitemOpen
  \bibfield  {author} {\bibinfo {author} {\bibfnamefont {J.~T.}\ \bibnamefont
  {Barreiro}}, \bibinfo {author} {\bibfnamefont {M.}~\bibnamefont
  {M\"{u}ller}}, \bibinfo {author} {\bibfnamefont {P.}~\bibnamefont
  {Schindler}}, \bibinfo {author} {\bibfnamefont {D.}~\bibnamefont {Nigg}},
  \bibinfo {author} {\bibfnamefont {T.}~\bibnamefont {Monz}}, \bibinfo {author}
  {\bibfnamefont {M.}~\bibnamefont {Chwalla}}, \bibinfo {author} {\bibfnamefont
  {M.}~\bibnamefont {Hennrich}}, \bibinfo {author} {\bibfnamefont {C.~F.}\
  \bibnamefont {Roos}}, \bibinfo {author} {\bibfnamefont {P.}~\bibnamefont
  {Zoller}}, \ and\ \bibinfo {author} {\bibfnamefont {R.}~\bibnamefont
  {Blatt}},\ }\href {\doibase 10.1038/nature09801} {\bibfield  {journal}
  {\bibinfo  {journal} {Nature}\ }\textbf {\bibinfo {volume} {470}},\ \bibinfo
  {pages} {486} (\bibinfo {year} {2011})}\BibitemShut {NoStop}%
\bibitem [{\citenamefont {Bosman}\ \emph {et~al.}(2017)\citenamefont {Bosman},
  \citenamefont {Gely}, \citenamefont {Singh}, \citenamefont {Bruno},
  \citenamefont {Bothner},\ and\ \citenamefont {Steele}}]{Bosman2017}%
  \BibitemOpen
  \bibfield  {author} {\bibinfo {author} {\bibfnamefont {S.~J.}\ \bibnamefont
  {Bosman}}, \bibinfo {author} {\bibfnamefont {M.~F.}\ \bibnamefont {Gely}},
  \bibinfo {author} {\bibfnamefont {V.}~\bibnamefont {Singh}}, \bibinfo
  {author} {\bibfnamefont {A.}~\bibnamefont {Bruno}}, \bibinfo {author}
  {\bibfnamefont {D.}~\bibnamefont {Bothner}}, \ and\ \bibinfo {author}
  {\bibfnamefont {G.~A.}\ \bibnamefont {Steele}},\ }\href {\doibase
  10.1038/s41534-017-0046-y} {\bibfield  {journal} {\bibinfo  {journal} {npj
  Quantum Inf.}\ }\textbf {\bibinfo {volume} {3}},\ \bibinfo {pages} {46}
  (\bibinfo {year} {2017})}\BibitemShut {NoStop}%
\bibitem [{\citenamefont {Thingna}\ \emph {et~al.}(2017)\citenamefont
  {Thingna}, \citenamefont {Barra},\ and\ \citenamefont
  {Esposito}}]{Thingna2017}%
  \BibitemOpen
  \bibfield  {author} {\bibinfo {author} {\bibfnamefont {J.}~\bibnamefont
  {Thingna}}, \bibinfo {author} {\bibfnamefont {F.}~\bibnamefont {Barra}}, \
  and\ \bibinfo {author} {\bibfnamefont {M.}~\bibnamefont {Esposito}},\ }\href
  {\doibase 10.1103/physreve.96.052132} {\bibfield  {journal} {\bibinfo
  {journal} {Phys. Rev. E}\ }\textbf {\bibinfo {volume} {96}},\ \bibinfo
  {pages} {052132} (\bibinfo {year} {2017})}\BibitemShut {NoStop}%
\bibitem [{\citenamefont {Tannor}(2007)}]{Tannor2007}%
  \BibitemOpen
  \bibfield  {author} {\bibinfo {author} {\bibfnamefont {D.~J.}\ \bibnamefont
  {Tannor}},\ }\href@noop {} {\emph {\bibinfo {title} {Introduction to Quantum
  Mechanics: A Time-Dependent Perspective}}}\ (\bibinfo  {publisher}
  {University Science Books},\ \bibinfo {year} {2007})\BibitemShut {NoStop}%
\bibitem [{\citenamefont {Zhu}\ \emph {et~al.}(2013)\citenamefont {Zhu},
  \citenamefont {Schmidt},\ and\ \citenamefont {Koch}}]{Zhu2013}%
  \BibitemOpen
  \bibfield  {author} {\bibinfo {author} {\bibfnamefont {G.}~\bibnamefont
  {Zhu}}, \bibinfo {author} {\bibfnamefont {S.}~\bibnamefont {Schmidt}}, \ and\
  \bibinfo {author} {\bibfnamefont {J.}~\bibnamefont {Koch}},\ }\href {\doibase
  10.1088/1367-2630/15/11/115002} {\bibfield  {journal} {\bibinfo  {journal}
  {New J. Phys.}\ }\textbf {\bibinfo {volume} {15}},\ \bibinfo {pages} {115002}
  (\bibinfo {year} {2013})}\BibitemShut {NoStop}%
\bibitem [{\citenamefont {Nicacio}\ \emph {et~al.}(2017)\citenamefont
  {Nicacio}, \citenamefont {Vald{\'{e}}s-Hern{\'{a}}ndez}, \citenamefont
  {Majtey},\ and\ \citenamefont {Toscano}}]{Nicacio2017}%
  \BibitemOpen
  \bibfield  {author} {\bibinfo {author} {\bibfnamefont {F.}~\bibnamefont
  {Nicacio}}, \bibinfo {author} {\bibfnamefont {A.}~\bibnamefont
  {Vald{\'{e}}s-Hern{\'{a}}ndez}}, \bibinfo {author} {\bibfnamefont {A.~P.}\
  \bibnamefont {Majtey}}, \ and\ \bibinfo {author} {\bibfnamefont
  {F.}~\bibnamefont {Toscano}},\ }\href {\doibase 10.1103/physreva.96.042341}
  {\bibfield  {journal} {\bibinfo  {journal} {Phys. Rev. A}\ }\textbf {\bibinfo
  {volume} {96}},\ \bibinfo {pages} {042341} (\bibinfo {year}
  {2017})}\BibitemShut {NoStop}%
\bibitem [{\citenamefont {de~Gosson}\ and\ \citenamefont
  {Nicacio}(2018)}]{Gosson2018}%
  \BibitemOpen
  \bibfield  {author} {\bibinfo {author} {\bibfnamefont {M.~A.}\ \bibnamefont
  {de~Gosson}}\ and\ \bibinfo {author} {\bibfnamefont {F.}~\bibnamefont
  {Nicacio}},\ }\href {\doibase 10.1063/1.5026586} {\bibfield  {journal}
  {\bibinfo  {journal} {J. Math. Phys.}\ }\textbf {\bibinfo {volume} {59}},\
  \bibinfo {pages} {052106} (\bibinfo {year} {2018})}\BibitemShut {NoStop}%
\bibitem [{\citenamefont {Wigner}(1932)}]{Wigner1932}%
  \BibitemOpen
  \bibfield  {author} {\bibinfo {author} {\bibfnamefont {E.}~\bibnamefont
  {Wigner}},\ }\href {\doibase 10.1103/physrev.40.749} {\bibfield  {journal}
  {\bibinfo  {journal} {Phys. Rev.}\ }\textbf {\bibinfo {volume} {40}},\
  \bibinfo {pages} {749} (\bibinfo {year} {1932})}\BibitemShut {NoStop}%
\end{thebibliography}%

\end{document}